\documentclass[11pt]{article}
\usepackage{graphicx,amssymb,amsmath}
\pdfoutput=1
\newcommand{\cC}{{\mathcal C}}
\newcommand{\cD}{{\mathcal D}}
\newcommand{\cE}{{\mathcal E}}
\newcommand{\cF}{{\mathcal F}}
\newcommand{\cG}{{\mathcal G}}
\newcommand{\cH}{{\mathcal H}}
\newcommand{\cJ}{{\mathcal J}}
\newcommand{\cL}{{\mathcal L}}
\newcommand{\cR}{{\mathcal R}}
\newcommand{\cX}{{\mathcal X}}
\newcommand{\cY}{{\mathcal Y}}
\newcommand{\cZ}{{\mathcal Z}}
\newcommand{\cDo}{\cD_{\rm out}}
\newcommand{\cDi}{\cD_{\rm in}}
\newcommand{\e}{\epsilon}
\newcommand{\Xs}{X_{\rm S}}
\newcommand{\Xmax}{X_{\rm max}}
\newcommand{\vc}{v_{\rm c}}
\newcommand{\R}{{\mathbb R}}
\DeclareMathOperator{\sech}{sech}
\DeclareMathOperator{\sign}{sign}
\DeclareMathOperator{\Imag}{Imag}
\DeclareMathOperator{\Real}{Real}
\DeclareMathOperator{\trace}{trace}
\newcommand{\w}{\omega}
\newcommand{\td}{t_{\rm delay}}
\newcommand{\subfig}[1]{\textbf{(#1)}}
\newcommand{\abs}[1]{\left\lvert #1\right\rvert}
\newtheorem{theorem}{Theorem}[section]

\title{Chaotic scattering in solitary wave interactions: A singular iterated-map description}\author{Roy Goodman%
\thanks{Department of Mathematical Sciences, New Jersey Institute of Technology, Newark, NJ 07102}}
\date{}
\begin{document}
\maketitle
\abstract{%
We derive a family of singular iterated maps---closely related to Poincar\'e maps---that describe chaotic interactions between colliding solitary waves.  The chaotic behavior of such solitary wave collisions depends on the transfer of energy to a secondary mode of oscillation, often an internal mode of the pulse.  Unlike previous analyses, this map allows one to understand the interactions in the case when this mode is excited prior to the first collision.  The map is derived using Melnikov integrals and matched asymptotic expansions and generalizes a ``multi-pulse'' Melnikov integral and allows one to find not only multipulse heteroclinic orbits, but exotic periodic orbits.  The family of maps derived exhibits singular behavior, including regions of infinite winding.  This problem is shown to be a singular version of the conservative Ikeda map from laser physics and connections are made with problems from celestial mechanics and fluid mechanics.
}

\textbf{%
Solitary waves are solutions to time-dependent partial differential equations (PDE) which can be described as profiles of unchanging spatial shape which move at constant velocity.  A fundamental question is what happens when two of them collide.  In certain classes of PDE a phenomenon known as chaotic scattering is seen in collisions--two waves will appear to bounce off each other several times and eventually move apart, with the number of ``bounces'' and the speed at which the waves eventually separate shown to be a very complicated function of the waves' initial speeds.  This has been seen in numerical simulations dating back twenty-five years but never fully explained mathematically.  We study a small system of ordinary differential equations, which may be derived as an approximation to the PDE dynamics.  This system has been shown to reproduce many of the features of the chaotic scattering and is much simpler to analyze.  We show that the simplified system may be further reduced into an even simpler set of equations called an iterated maps.  These maps are then analyzed using the tools of dynamical systems.  We  find significant mathematical structure, including many bifurcations and ``infinite horseshoes'', in the iterated map that are responsible for chaotic scattering.
}

\section{Introduction}
\label{sec:intro}
Solitary waves---localized solutions to partial differential equations (PDEs) which translate at uniform velocity with a constant spatial profile---are a ubiquitous phenomenon in physical sciences.  A fundamental question relating to these objects is their behavior upon collisions, either with other solitary waves, or with localized changes (``defects'') to the medium through which they propagate.  

The dynamics of such collisions depends, of course, on the PDE in question, but there exist classes of equations in which qualitatively similar phenomena are expected.  In strongly dissipative systems, two colliding solitary waves generally lose their distinct identities and merge into a single larger bump.  At the opposite extreme are completely integrable systems whose solitary waves (solitons) survive collisions with their identities intact, due to a linear structure hidden deep in the underlying equations. This was first seen numerically by Zabusky and Kruskal~\cite{ZabKru:65} and confirmed by the discovery of the exact two-soliton solution~\cite{Lax:68}, both for the Korteweg-de Vries equation.

An interesting case where the spectrum of possible behaviors is much richer is that of systems which are both \emph{non}-dissipative and \emph{non}-integrable.  Chaotic scattering between colliding solitary waves is a problem that has been rediscovered by numerous groups since first being hinted at numerical simulations in the 1970's~\cite{AblKruLad:79}.  We defer a discussion of a history of the problem to the next section.

\begin{figure}
\begin{center}
\includegraphics[width=1.5in]{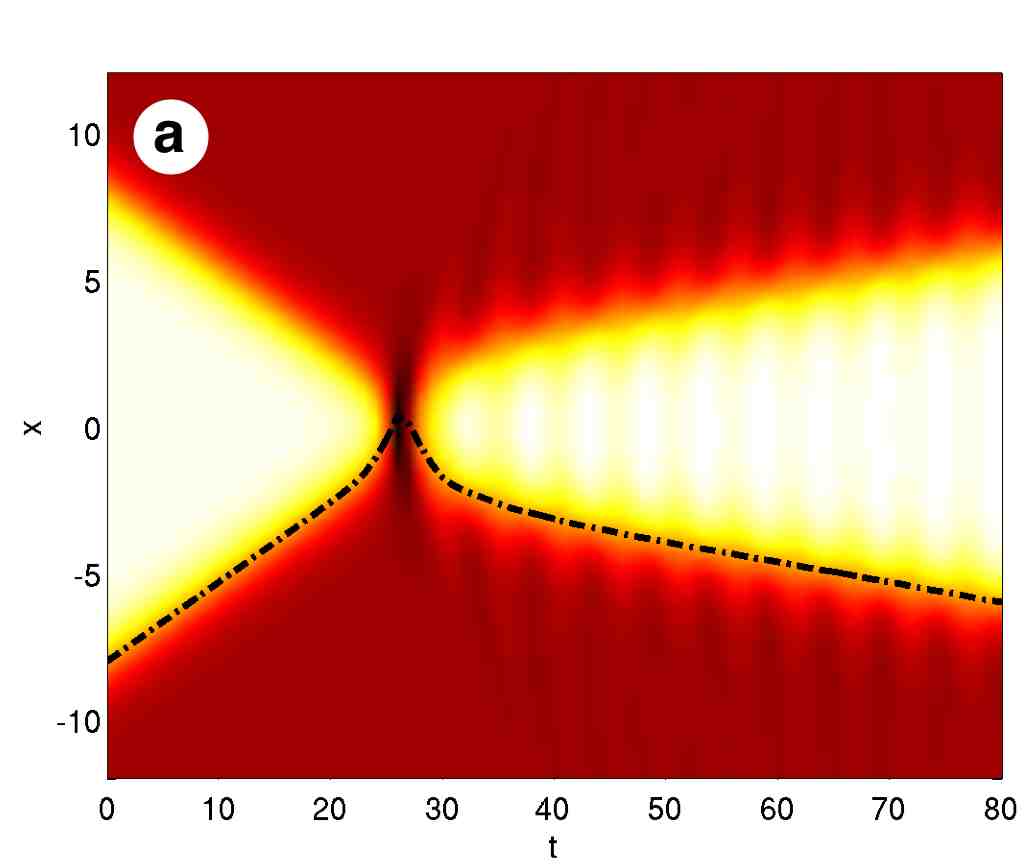}
\includegraphics[width=1.5in]{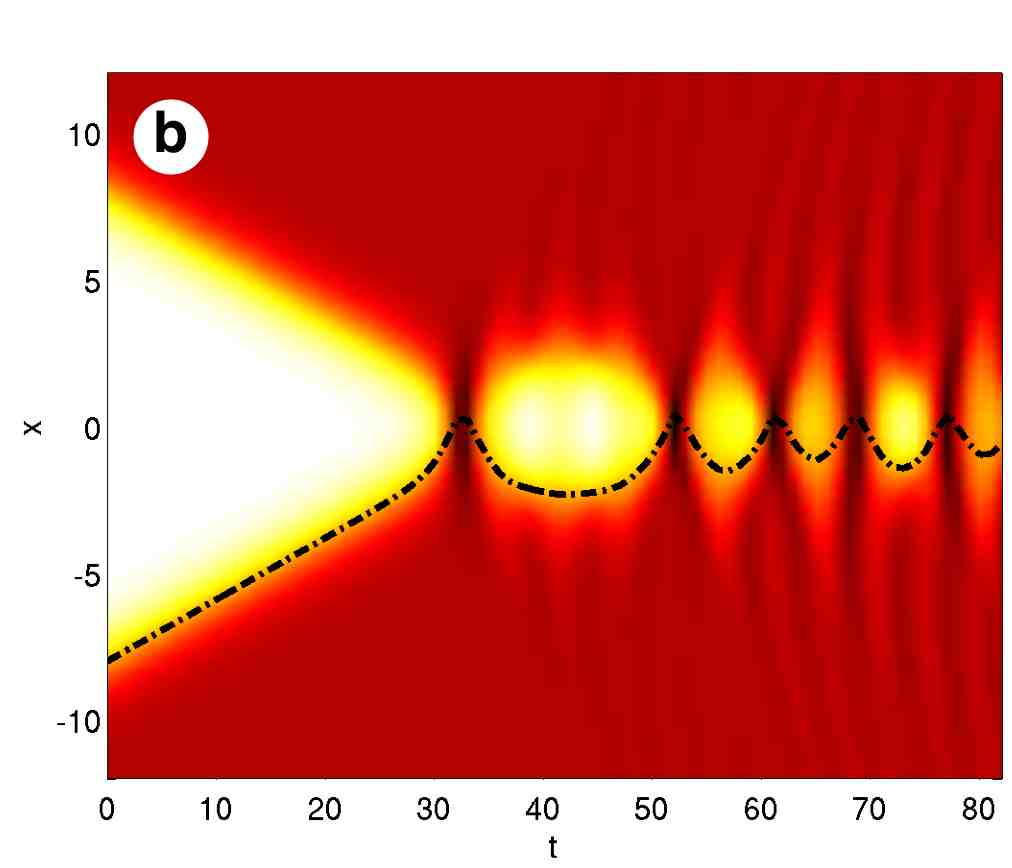}
\includegraphics[width=1.5in]{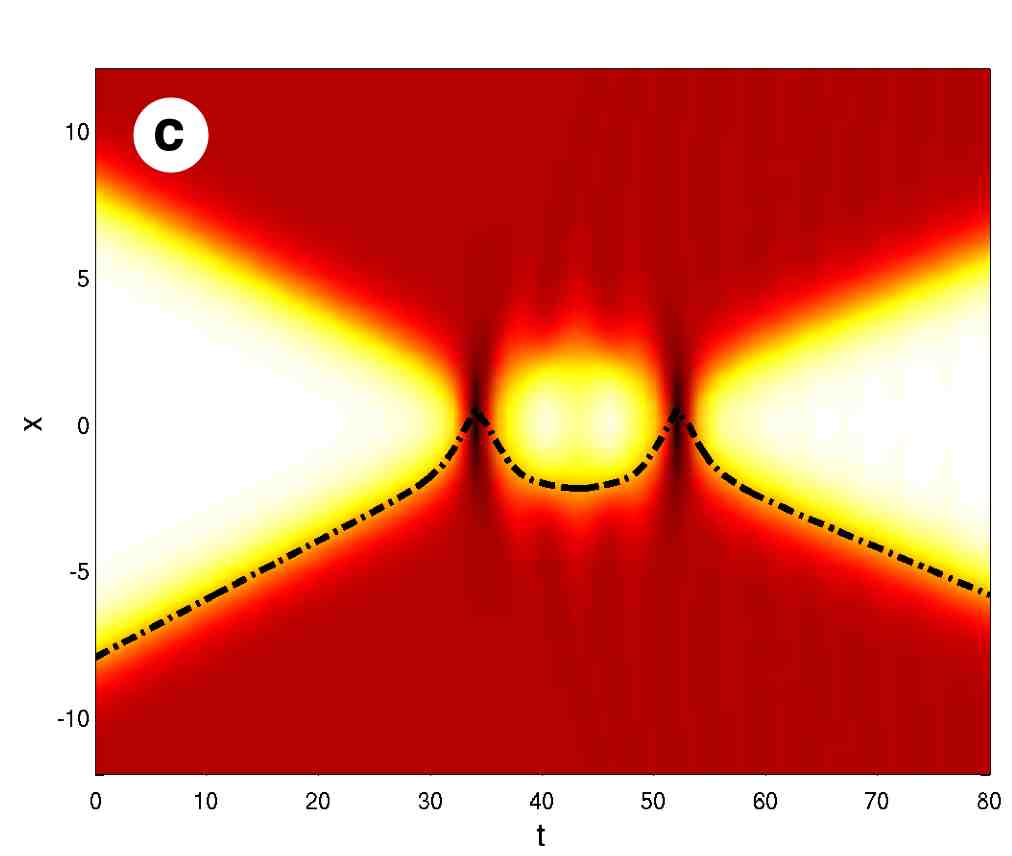}
\includegraphics[width=1.5in]{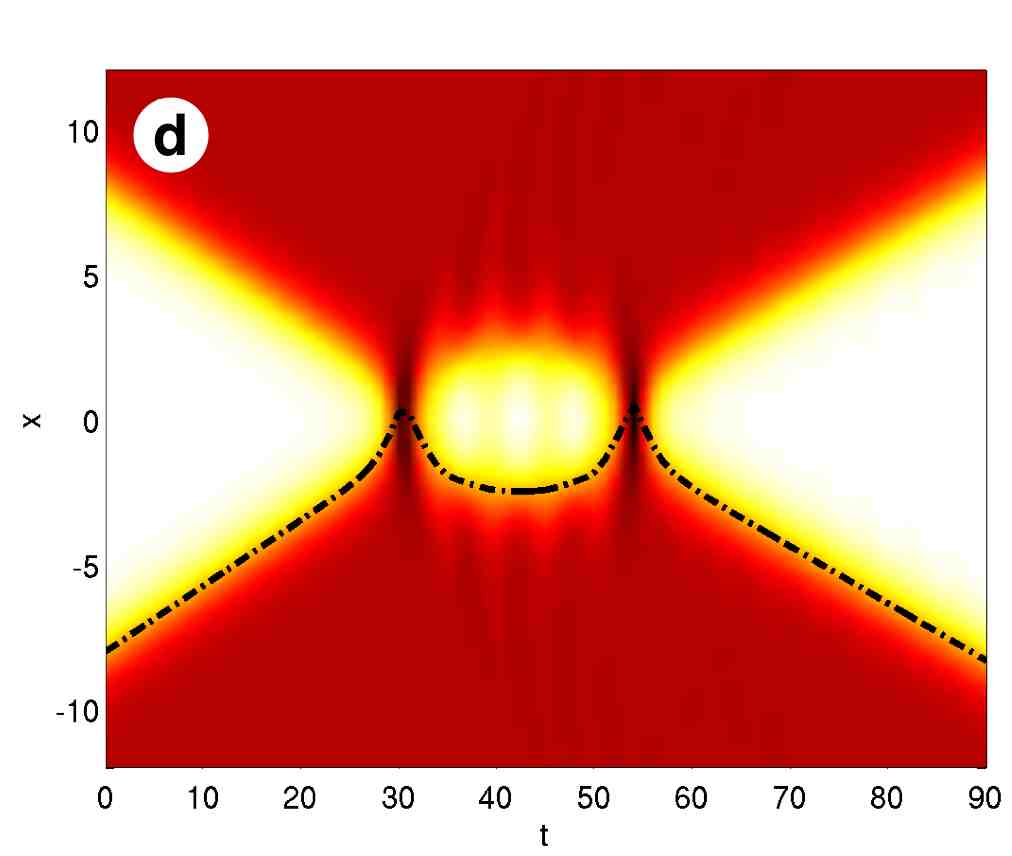}
\includegraphics[width=1.5in]{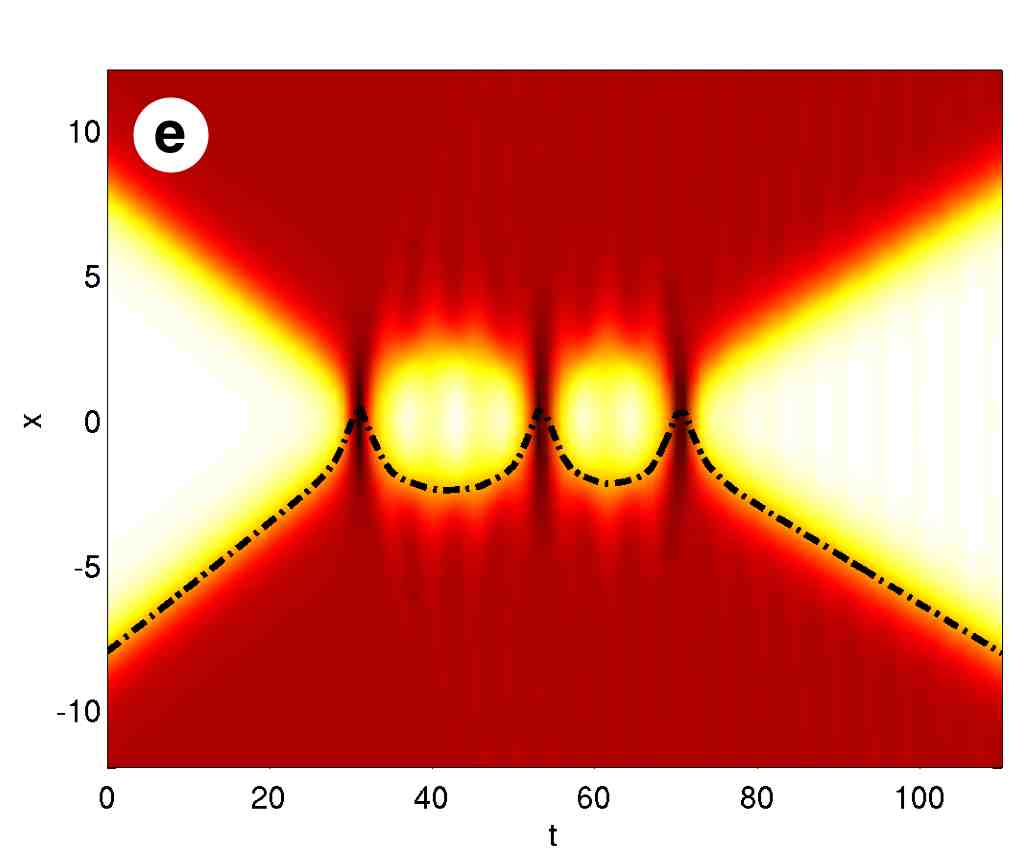}
\caption{(Color online) Kink-antikink collisions in~\eqref{eq:phi4}. (a) 1-bounce solution at $v= 0.27 >\vc$  (b) capture at $v=0.21$, (c) and (d) 2-bounce solutions at $v=0.1987$ and $v=0.2268$, (e) 3-bounce solution at $v=0.2228$. The position of one kink, determined by fitting the numerical solution to a kink of undetermined position, is given by the dashed black lines.}
\label{fig:phi4}
\end{center}
\end{figure}

We consider the particular example of the $\phi^4$ equation which arises as a model problem in many areas of theoretical physics:
\begin{equation}
\label{eq:phi4}
\phi_{tt}-\phi_{xx} -\phi + \phi^3 = 0.
\end{equation}
This supports ``kink'' solitons of the form
\begin{equation}
\label{eq:kink}
\phi(x,t)= \pm \tanh{\frac{\xi}{\sqrt{2}}},
\end{equation}
where $\xi=\frac{x-x_0-vt}{\sqrt{1-v^2}}$ and the kink velocity may take any value $\abs{v}<1$, and the choice of a minus sign in the above formula describes the so-called antikink.    Following~\cite{AblKruLad:79,AOM:91,CSW:83}, we numerically simulate the collision of a kink-antikink pair with equal and opposite velocity, and show the results in figure~\ref{fig:phi4}.

\begin{figure}
\begin{center}
\includegraphics[width=3in]{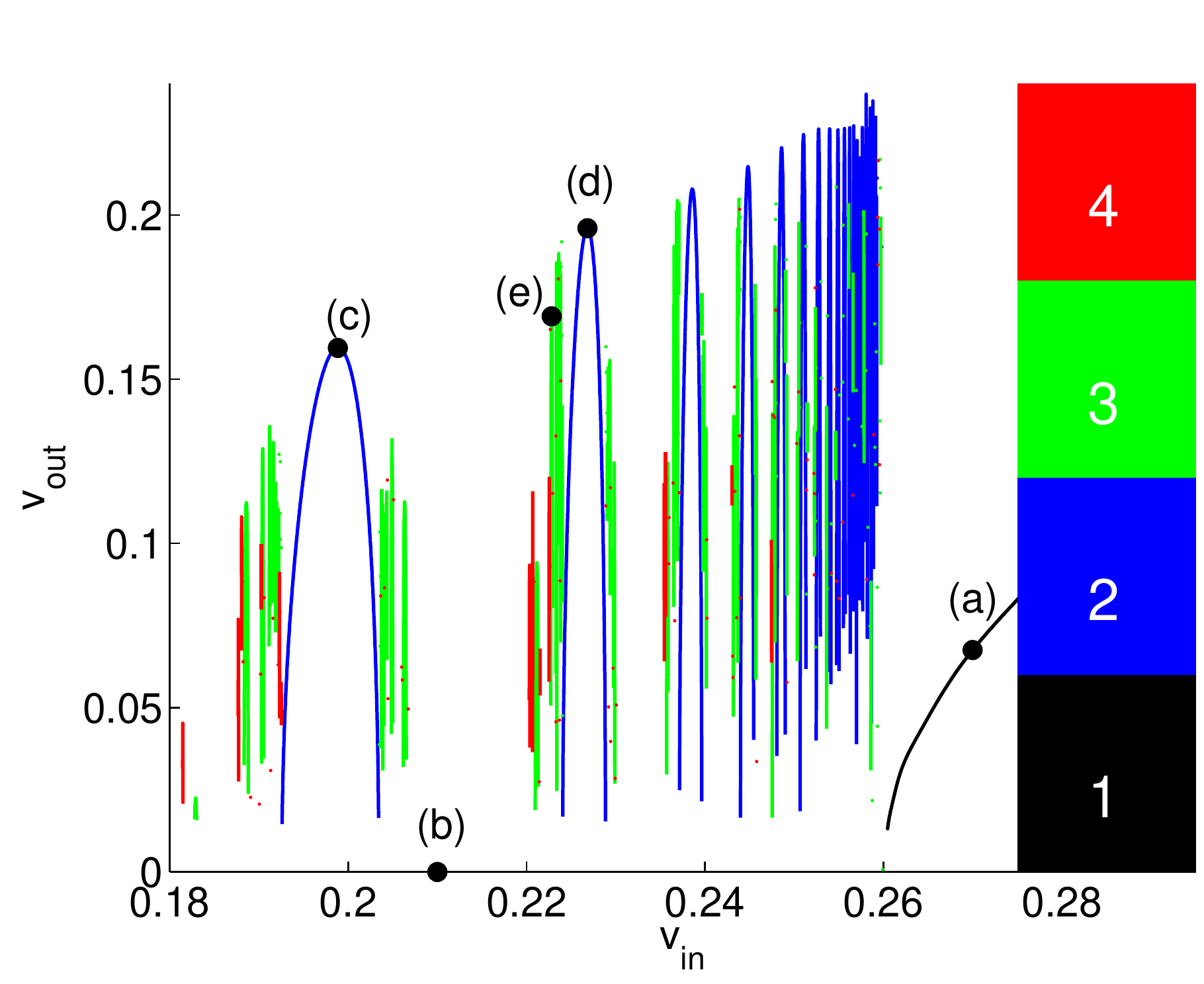}
\caption{(Color online) The output velocity as a function of the input velocity for kink-antikink collisions, color coded by number of times the kink and antikink ``bounce'' off each other before escaping.  The labels (a)-(e) give the corresponding solutions in figure~\ref{fig:phi4}.}
\label{fig:phi4v}
\end{center}
\end{figure}

For waves with initial speeds above some critical velocity $\vc$ the waves separate, albeit with reduced speed, figure~\ref{fig:phi4}a.  For most initial velicities below $\vc$, the waves are captured and form into a single localized structure, figure~\ref{fig:phi4}b.  For initial velocities in certain ``resonance windows,'' however, they may eventually separate, as in figure~\ref{fig:phi4}c, d, and e.  
Plots 1c and 1d show `two bounce' solutions, as one can see from the dashed line that the kinks collide twice.  One may discern that the transient bound state in figure~\ref{fig:phi4}c oscillates twice (the brightest white spots in the middle) while the transient bound state in figure~\ref{fig:phi4}d oscillates three times.  Figure~\ref{fig:phi4}e shows a 3 bounce solution. In figure~\ref{fig:phi4v}, we show how the escape velocity $v_{\rm out}$ depends on the initial velocity~$v_{\rm in}$. The initial conditions corresponding to the 5 experiments in figure 1 are marked in figure~\ref{fig:phi4v}. Also plotted, using color, is the number of times the kink and antikink ``bounce off'' each other before escaping.  In fact, we have found significant detail beyond what can be reasonably depicted in this figure.  The structure is repeated at smaller widths with larger numbers of bounces before escape, and we have found numerical solutions with as many as nine bounces.  Of course, as this PDE displays sensitive dependence on initial conditions, it also has sensitive dependence on numerical discretization and the fractal structure persists, but shifts slightly with small changes to the discretization.  From figure~\ref{fig:phi4v}, it is clear that the behavior of these colliding solitary waves is an example of chaotic scattering; see~\cite{OttTel:93} for an introduction to a special issue dedicated to this topic.

\begin{figure}
\begin{center}
\includegraphics[height=2in]{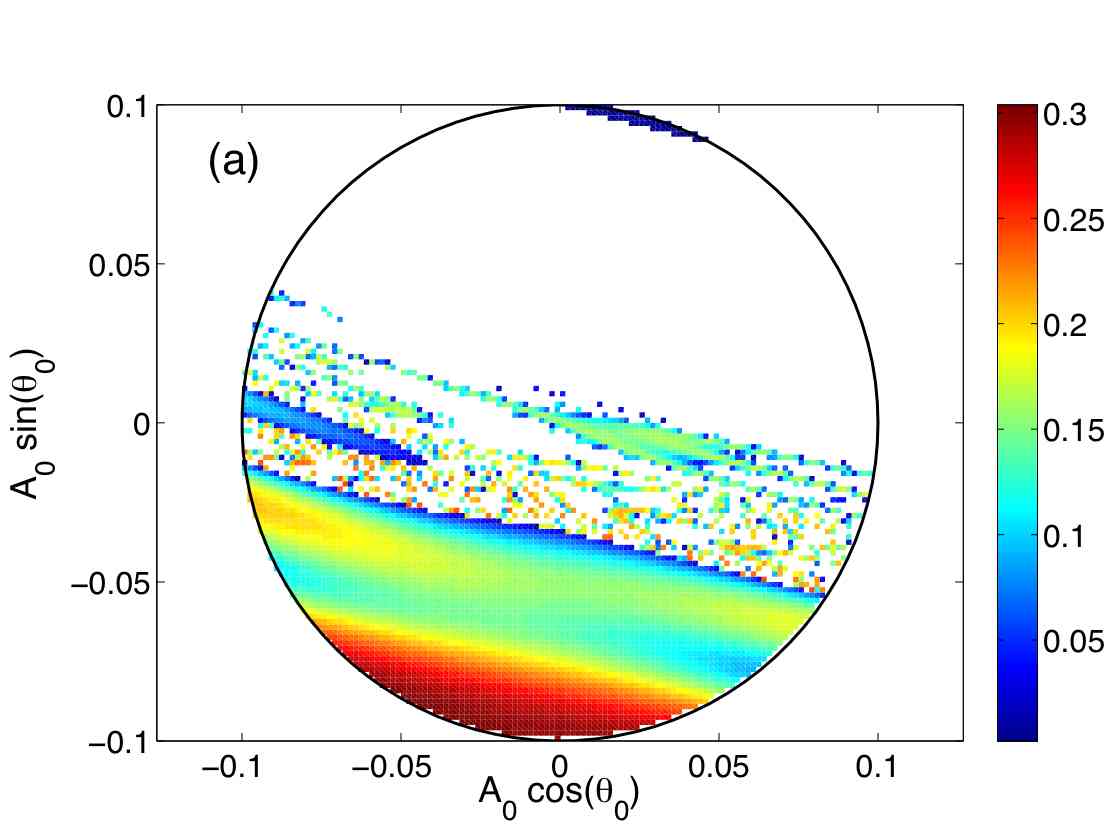}
\includegraphics[height=2in]{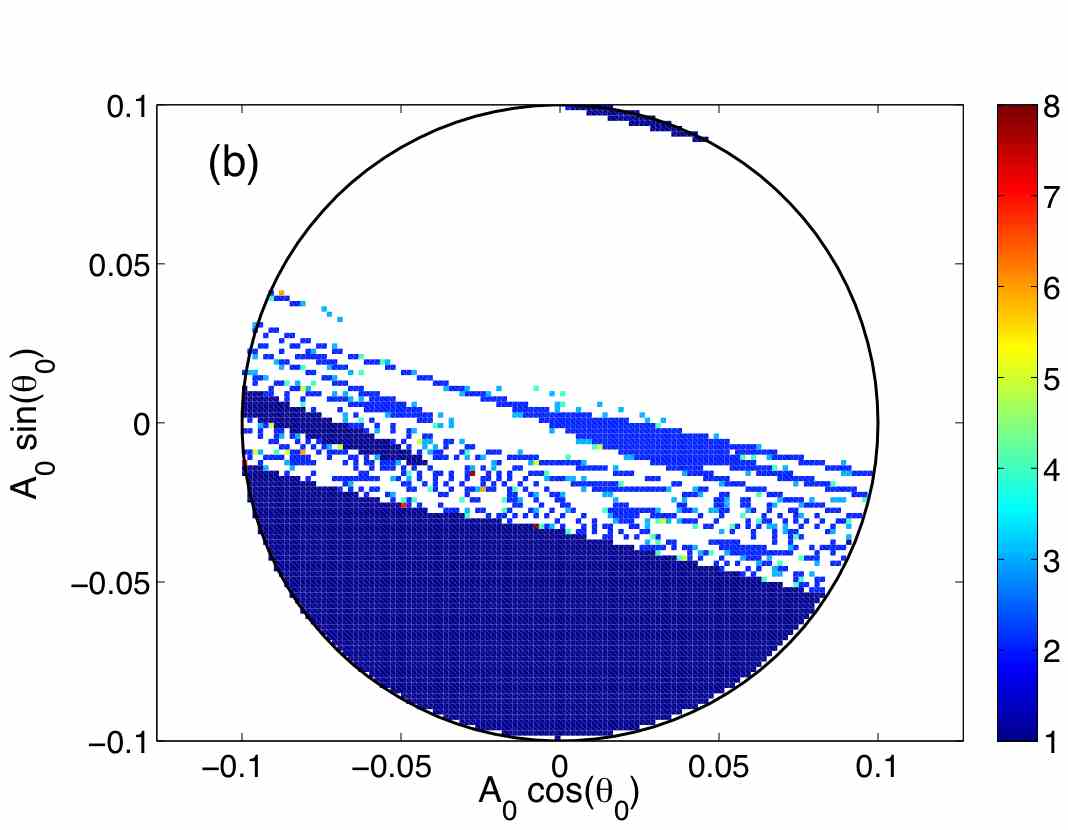}
\caption{(Color online) For this figure a kink and antikink were initialized with incoming speed $v_{\rm in}=0.19871$, the center of the first two-bounce resonance in figure~\ref{fig:phi4v}, and an internal mode pre-excited with parameters $A_0$ and $\theta_0$.  \subfig{a}, the speed with which the kink and antikink separate;~\subfig{b}, the number of times the kink and antikink collide before escaping.}
\label{fig:phi4_excited}
\end{center}
\end{figure}

Notice in figure~\ref{fig:phi4} that the escaping solution has an additional oscillation not present before the two solutions collide.  In fact this oscillation is evidence of an internal mode which is excited during the collision, and which is responsible, as we will discuss below, for both the capture dynamics when $v<\vc$ and the subsequent escape in the case of resonance.  Linearizing the $\phi^4$ equation~\eqref{eq:phi4} about the kink~\eqref{eq:kink}, and looking for solutions of the form $\chi_1(\xi) \cos{\w_1 \tau}$, where $\tau= (t-v (x-x0))/\sqrt{1-v^2}$ is time in the Lorenz-shifted coordinate frame, one finds solutions
\begin{equation}
\label{eq:chi1}
\chi_1(\xi)= \bigl(\frac {3} {\sqrt{2}} \bigr)^{1/2}\tanh{\frac {\xi} {\sqrt{2}}} \sech{\frac {\xi} {\sqrt{2}}} 
\end{equation}
with frequency $\w_1=\sqrt{\frac 3 2}$. 
A question of interest, then, is what happens if the kink's internal mode is excited before the collision.  This is depicted in figure~\ref{fig:phi4_excited}.  The collision is initialized with 
$$\phi(x,0) = \phi_0 (x;x_0, \theta_0, A_0) - \phi_0 (x;-x_0, \theta_0, A_0)-1,$$
where
$$
\phi_0(x;x_0,\theta_0,A_0) = \phi_K(\xi) + A_0 \chi_1(\xi)\cos{(\w_1 \tau-\theta_0)}.
$$
The antikink and its internal mode are exact mirror images of the kink and its internal mode, and equation~\eqref{eq:phi4} preserves even symmetry.  Each of the approximately 11500 pixels represents a numerical solution of~\eqref{eq:phi4}. We did not consider the more complicated case where the kink, antikink, and their internal modes are chosen without this symmetry.  The methods developed by Goodman and Haberman in~\cite{GH:04,GooHab:05,GH:05,GooHab:07} are sufficient to describe much of the behavior depicted in figures~\ref{fig:phi4} and~\ref{fig:phi4v}, although the form of the map derived in the current paper for the more general class of initial conditions allows a much fuller understanding of the dynamics, especially of the structures seen in figure~\ref{fig:phi4_excited}.

The goal of this paper is to explain in detail the mechanisms underlying the chaotic scattering behavior described above and in particular to describe how to frame this system in such a way that the standard techniques of dynamical systems theory may be applied.  We will proceed as follows.  Section~\ref{sec:prelim} contains further preliminaries including a historical overview of the phenomena, and an introduction of  ``collective coordinate'' ordinary differential equation (ODE) systems used to model the PDE dynamics. 
In section~\ref{sec:derivation} we derive, in detail, an iterated map that functions as a sort of Poincar\'e map for the ODE model.  We then interpret the map and show the results of direct numerical simulations in section~\ref{sec:interp}.  In section~\ref{sec:morse}, we will analyze its dynamics.    In section~\ref{sec:extensions}, we extend the analysis to two related systems, one from a different solitary-wave collision problem and the second from geophysical fluid mechanics and find interesting and more complex iterated maps.We conclude in section~\ref{sec:conclusion} with a discussion of how dissipation---caused in the PDE by the irreversible loss of energy to escaping radiation---may be introduced into the maps and used to explain capture.

\section{Further Preliminaries}
\label{sec:prelim}
\subsection{Background and technological motivation}
We have summarized the history of this problem in our previous work; see especially~\cite{GooHab:07}. The two bounce resonance was discovered and explored in a series of papers by Campbell, Peyrard, and various collaborators in the 1980's~\cite{CPS:86,CP:86,CSW:83,PC:83,RemPey:84,MalCamKno:93} following some hints in earlier numerical experiments by Ablowitz et al.~\cite{AblKruLad:79}.  Campbell et al.\ showed via careful examination of numerical experiments and via heuristic arguments that the capture and subsequent escape of the solitary waves was due to resonant energy transfer to a secondary mode, namely the kink's internal mode.  Their calculations  for the locations of the two-bounce windows, based on parameter-fitting to their numerical simulations, are remarkably accurate but do not give any insight into how to find $\vc$ or how the window locations might depend on parameters in the equations. The fractal structure was studied in somewhat more detail by Anninos et al~\cite{AOM:91}.  

The two-bounce resonance phenomenon was subsequently observed by Fei, Kivshar, and V\'azquez in the collisions of kink-like solitary waves with localized defects~\cite{FKV:92,FKV:92a}, and then by Tan and Yang in the collisions between solitary waves in a system of coupled nonlinear Schr\"odinger equations describing light propagation in birefringent optical fibers~\cite{TY:01,TY00}. More recently it has been seen in quantum field theory in the interaction of topological solitons with defects in the metric of the background spacetime~\cite{Jav:06,PieZak:07}.  In section~\ref{sec:extensions} we will extend our methods to the system studied by Fei et al.\ in~\cite{FKV:92}.

There have been a few studies that looked at the behavior of solitary waves when the secondary mode is excited prior to the collision, for example in Fei et al.~\cite{FKV:92} and Forinash et al~\cite{FPM:94}.  These have been mainly small-scale numerical simulations. One goal of this paper is to investigate such collisions more thoroughly.

Many previous studies have used collective-coordinate models to study collision phenomena.    In such models, the infinite-dimensional  dynamics are reduced to a finite-dimensional system of ordinary differential equations (ODE).  These ODE models are derived via the ``variational method,' a non-rigorous procedure based on the underlying Lagrangian structure of the system.  Such methods are reviewed by Malomed~\cite{Mal:02}.  In recent work with Haberman, we have used these collective coordinate models to derive approximate formulas for the critical velocity and derived simple iterated maps that reproduce the fractal structure of figure~\ref{fig:phi4v}.  In section~\ref{sec:derivation} we derive a more general form of this map which applies to a wider class of initial conditions and is more amenable to the methods of dynamical systems.

In the earlier of the above-cited works, the PDE simulations typically used finite-difference (second or fourth order) discretization of the spatial derivatives and explicit Runge-Kutta time stepping.  Figures~\ref{fig:phi4} and~\ref{fig:phi4v} were produced using pseudospectral (cosine-transform) spatial discretizations and time-stepping algorithms which treat the (stiff) linear portion of the equation exactly while applying an explicit fourth-order Runge-Kutta time step to the nonlinear terms~\cite{KasTre:05}.  This allows for the use of a coarser spatial mesh and much larger time steps, and, more importantly, eliminates the discretization-induced damping which may have made it much harder to resolve the narrow windows of chaotic scattering.

One motivation that has often been cited for studying solitary wave collisions, especially collisions with defects, is the desire to build all-optical communications systems.  Current technologies send signals at high speeds using pulses of light.  Such signals are converted to electronic form for processing and then converted back to optical form for retransmission.  The idea is that one may engineer the nonlinear effects in an optical medium to allow the data processing and thus avoid the time-consuming conversion back and forth to electronics.
In one such scenario for an optical memory, a defect in an optical medium might be used to ``trap'' a pulse of light~\cite{AceDoh:06,GSW}. At a later time, a second ``probe'' pulse might be sent in to detect whether a pulse has previously been captured.  The computation depicted in figure~\ref{fig:phi4_excited} is meant to give an idea of the possible behavior of such a system.  This may be seen as either a blessing or a curse: a curse because to ensure that a chaotic scatterer gives the desired output for a given input requires very accurate control on the input, and a blessing because, once obtaining such precision, there are a large number of possible outputs that could be used to encode different pieces of information.

\subsection{Collective Coordinate ODE Models}
Solutions to the $\phi^4$ equation minimize the action with Lagrangian density
\begin{equation}
\label{eq:pde_lagrangian}
{\mathcal L}(\phi) = \frac{1}{2} \phi_t^2 -  \frac{1}{2} \phi_x^2 + \frac{1}{2} \phi^2  -\frac{1}{4} \phi^4.
\end{equation}
A system of ODEs that models the behavior of the kink-antikink collision is derived using the ``variational approximation.''  Instead of looking for minimizers of 
\begin{equation}
\label{eq:action}
\iint \cL(\phi)\ dx\ dt,
\end{equation} we look for minimizers among functions of a predetermined spatial form dependent on a finite number of time-dependent parameters.  For this particular problem, there are two such parameters: $X(t)$ describing the separation between the kink and antikink, and $A(t)$ giving the amplitude of the internal mode $\chi_1$.  This parameter-dependent profile is substituted into~\eqref{eq:action}, and the inner $x$ integral is evaluated explicitly, giving an action dependent only on $t$.  Computing the Fr\'echet derivative of this action gives the Euler-Lagrange equations for the evolution of $X(t)$ and $A(t)$,
\begin{subequations}
\label{eq:AOM}
\begin{align}
\ddot X &= \frac{1}{2+2I(X)}\left(-I'(X)\dot X^2 -U'(X)+  F(X)'A \right) \label{eq:AOM_X}\\
\ddot A +\w_1^2 A &=F(X).\label{eq:AOM_A}
\end{align}
\end{subequations}

\begin{figure}
\begin{center}
\includegraphics[height=2in]{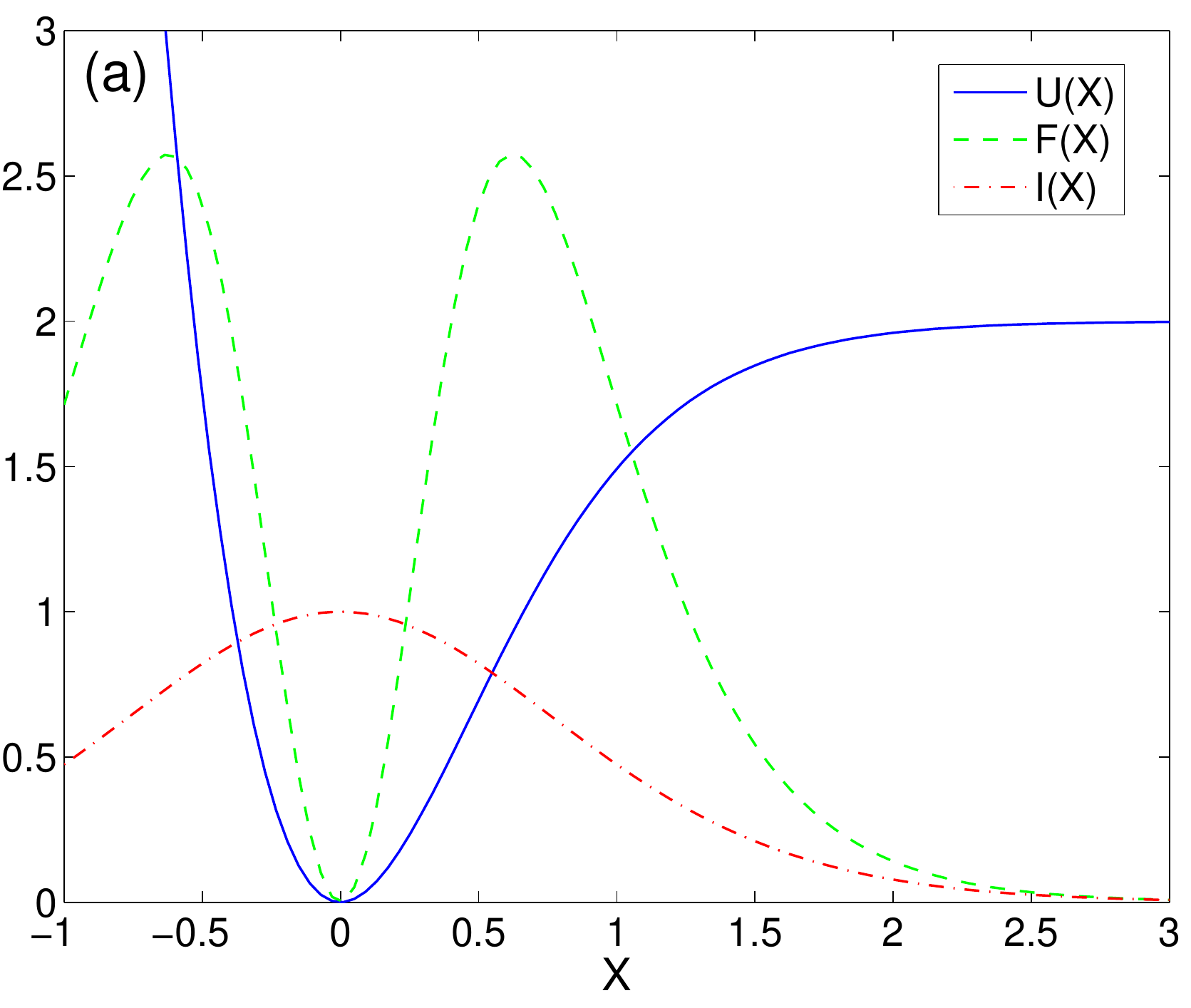}
\includegraphics[height=2in]{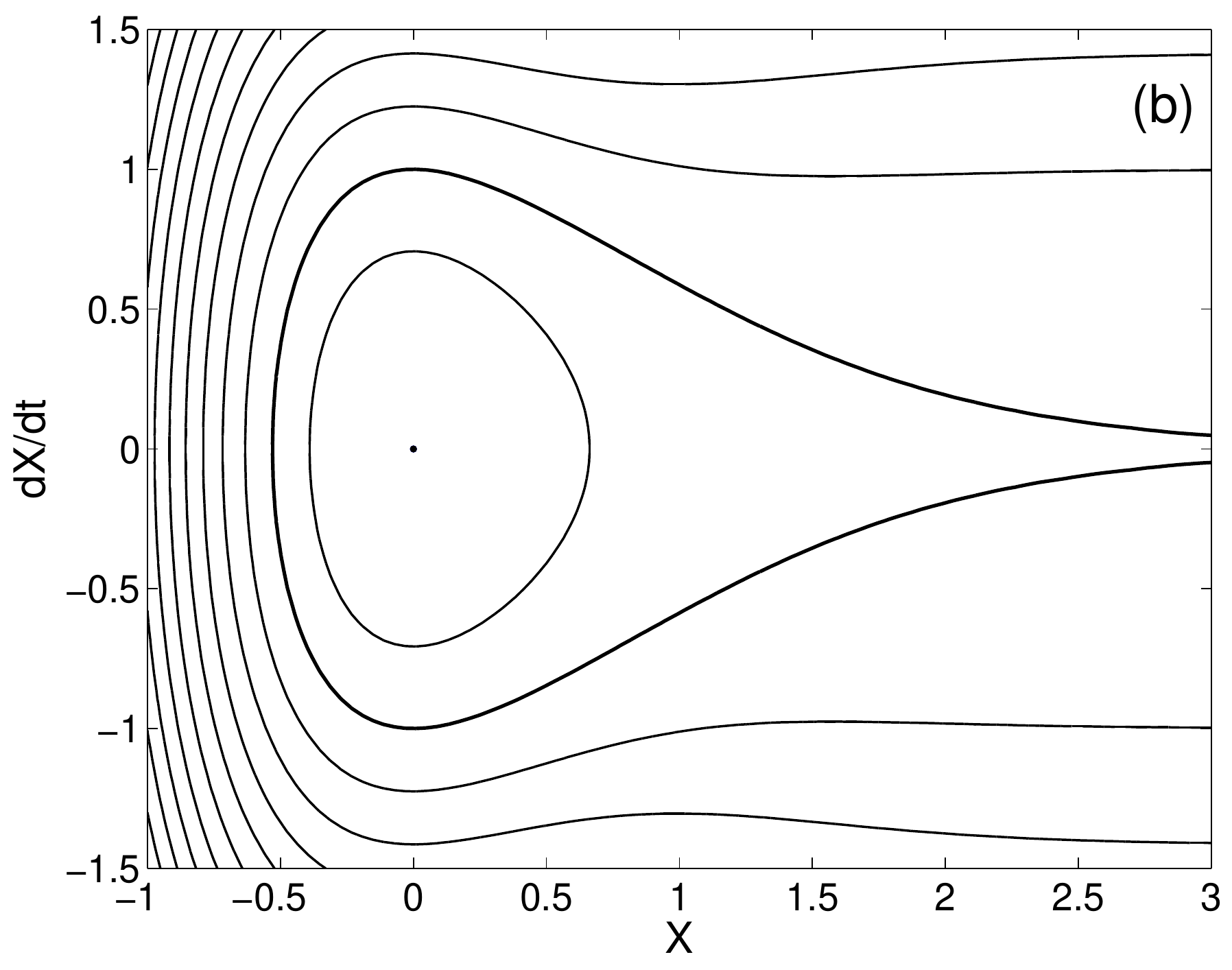}
\caption{(Color online) \subfig{a}The functions defining system~\eqref{eq:AOM}. \subfig{b} The phase-plane for equation~\eqref{eq:AOM_X}, with the coupling term $F'(X)A$ ignored.}
\label{fig:AOM}
\end{center}
\end{figure}

We will study a simplification of these equations, so the exact forms of all the terms will not be important. The relevant details are summarized in figure~\ref{fig:AOM}a.  The terms $I(X)$ (position-dependence of mass) and $F(X)$ (coupling) decay exponentially for large $\abs{X}$, while the potential $U(X)$ has a single minimum and unbounded for $X<0$ and decays exponentially to $U=2$ as $X\to\infty$.  Ignoring the coupling term $F'(X)A$ in~\eqref{eq:AOM_X}, the uncoupled $X$-dynamics conserves an energy $E$, whose level sets give the trajectories  seen in figure~\ref{fig:AOM}b.  This system was shown in simulations by Anninos et al.\ to reproduce the chaotic scattering seen in the PDE system~\cite{AOM:91} and in~\cite{GooHab:05}, we showed how to determine the critical velocity and derived a restricted form of an iterated map that reproduces this behavior.

Here, as in~\cite{GooHab:07}, we simplify~\eqref{eq:AOM} by eliminating $I(X)$ and replacing $U(X)$ and $F(X)$ with simpler functions that preserve the main features of~\eqref{eq:AOM}, namely the topology of the uncoupled phase plane.  We have made this choice in order to clarify the exposition and will discuss later, where appropriate, how this effects the calculation. Our simplified model is
\begin{subequations}
\label{eq:morse}
\begin{align}
m \ddot X + U'(X) + \e F'(X) A  &=0 \label{eq:X_morse} \\
\ddot A + \w^2 A + \epsilon F(X) &=0,
\label{eq:A_morse}
\end{align}
\end{subequations}
where $$U(X) =e^{-2X}-e^{-X} \text{ and } F(X) = e^{-X}.$$
$U(X)$ is the potential for the Morse oscillator, a model for the electric potential of a diatomic model, used in quantum mechanics with exponentially weak long-distance interactions.  We note that this system conserves an energy (Hamiltonian) of the form
\begin{equation}
\label{eq:Hmorse}
H= \frac{m}{2}\dot X^2 + U(X) + \frac{1}{2}(\dot A^2 + \w^2 A^2) + \e F(X) A.
\end{equation}  

\begin{figure}
\begin{center}
\includegraphics[width=0.45\textwidth]{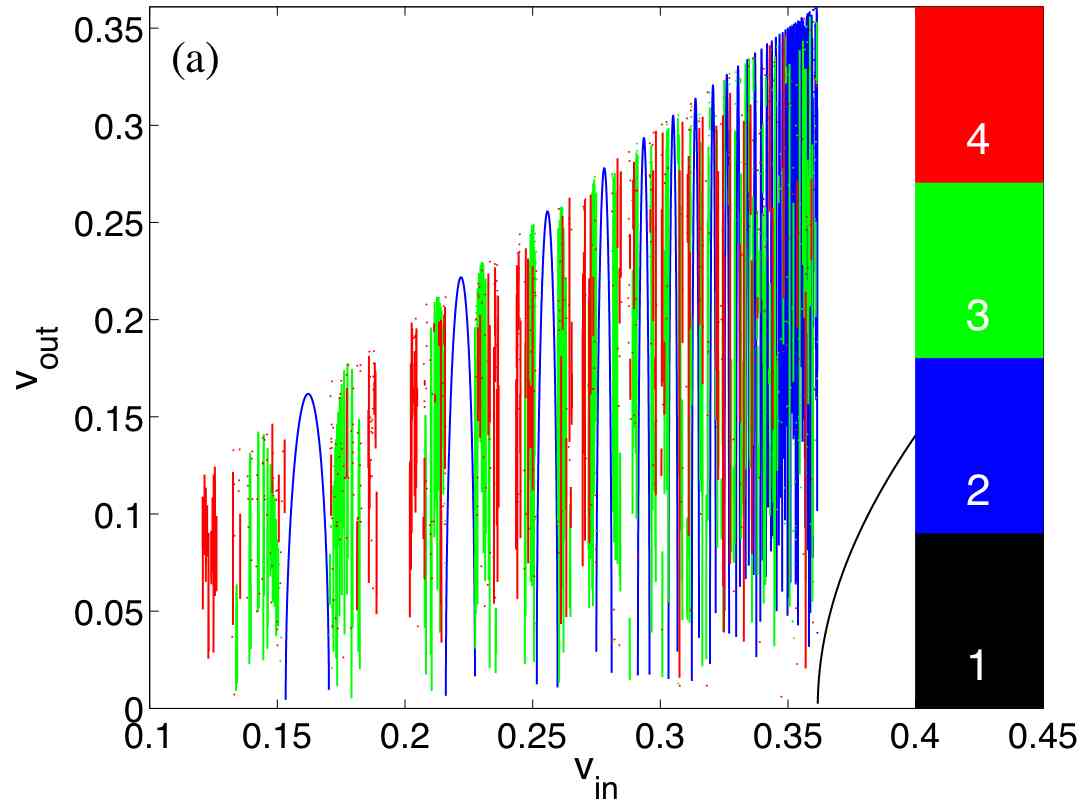}
\includegraphics[width=0.45\textwidth]{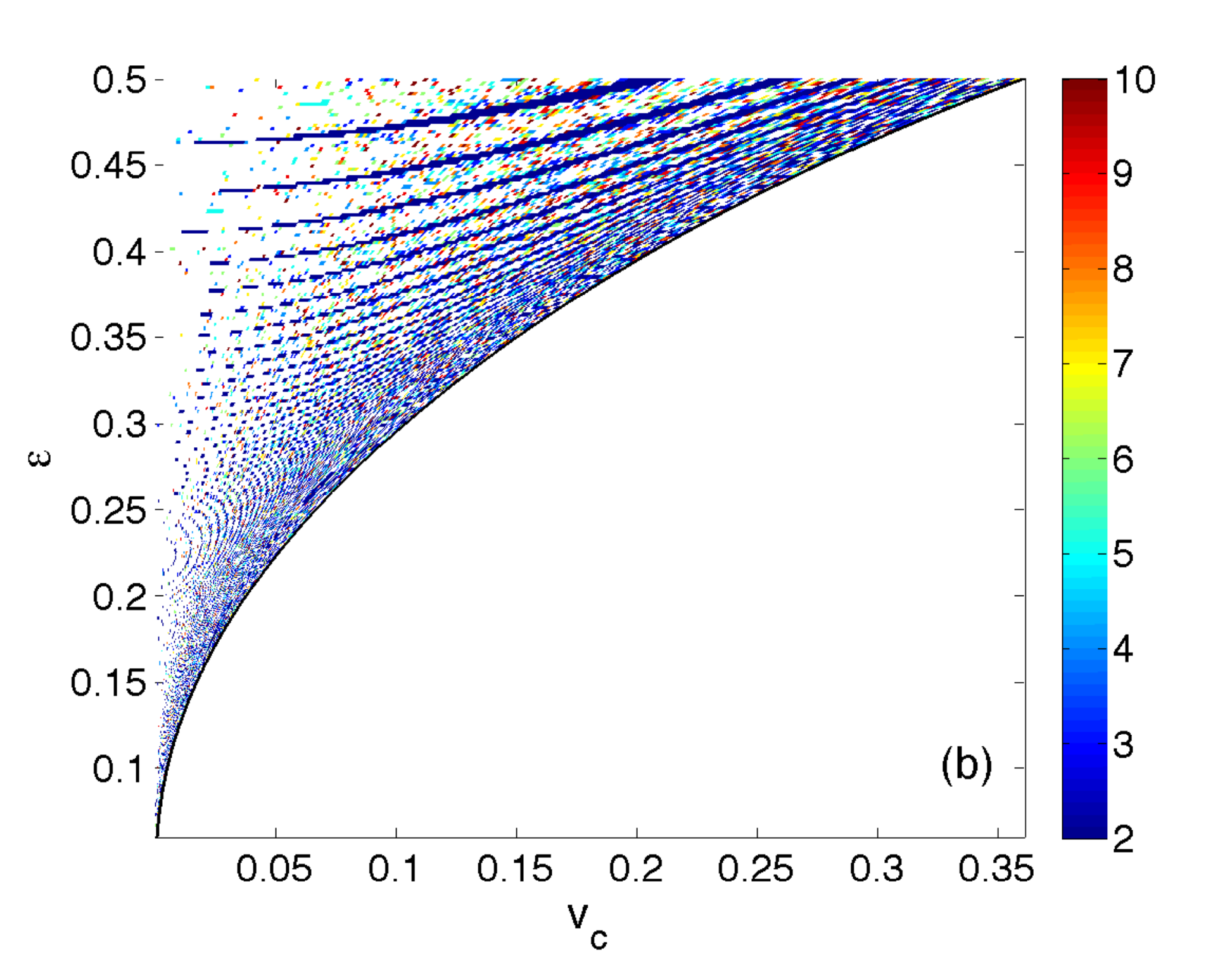}
\caption{(Color online) \subfig{a} The input vs.\ output velocity graph for~\eqref{eq:morse} with $m=\epsilon=1$ and $\w=2$. \subfig{b} The number of interactions preceding escape as a function of both $\epsilon$ and $v_{\rm in}$ with $m=1$ and $\w=\epsilon^{-1/2}$. }
\label{fig:ode_in_out}
\end{center}
\end{figure}

That such an ODE can reproduce the qualitative and, to a surprising extent, the quantitative features of the kink-antikink collisions has been well-documented~\cite{AOM:91,FKV:92a,FKV:92,TY:01,TY00}. Figure~\ref{fig:ode_in_out}(a) is the equivalent of figure~\ref{fig:phi4v} for equation~\eqref{eq:morse}, with impressive qualitative agreement.  A major difference between these two computations is that in the ODE model, capture only happens to a set of initial conditions of measure zero by conservation of phase-space volume, while in the PDE, a nonzero fraction of the initial conditions lead to capture.  Each of the windows has a finite width, although most in this figure are unresolved.  Figure~\ref{fig:ode_in_out}(b) shows, as a function of both $v$ and $\epsilon$, the number of interactions before the solution escapes to infinity.  Figure~\ref{fig:ode_in_out}(a) of the figure may be thought of as a horizontal slice through Figure~\ref{fig:ode_in_out}(b).

\begin{figure}
\begin{center}
\includegraphics[width=0.45\textwidth]{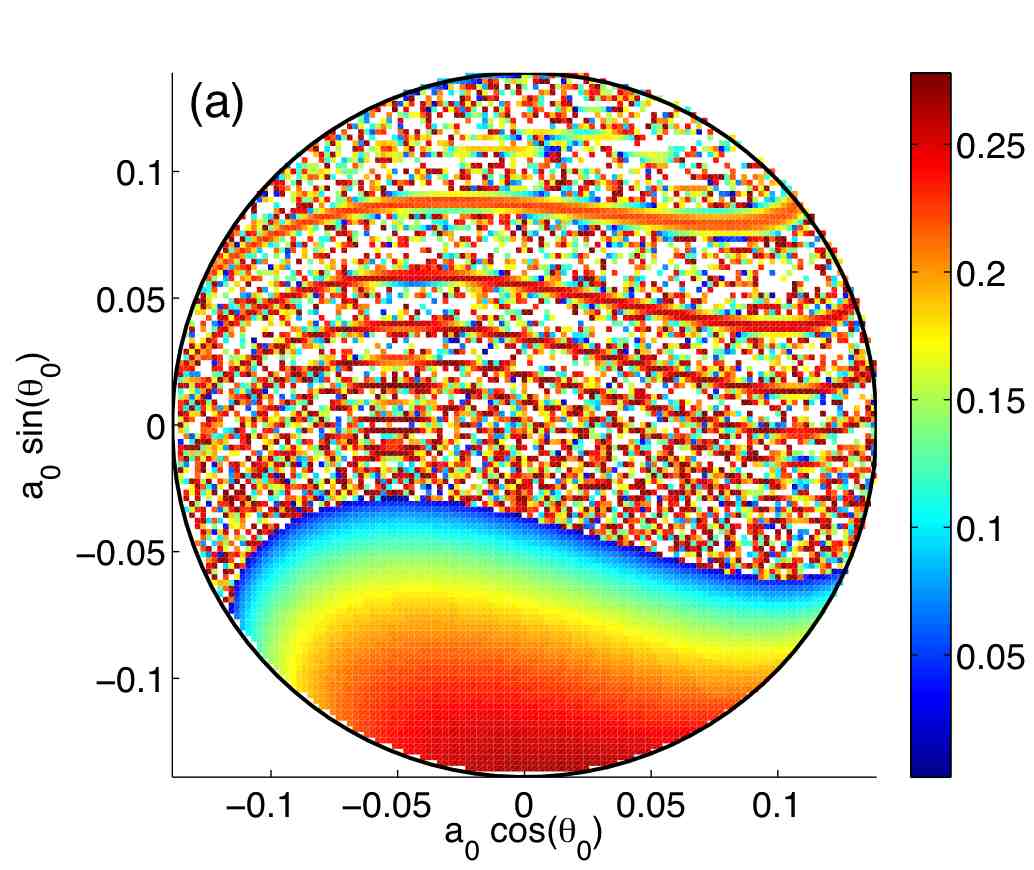}
\includegraphics[width=0.45\textwidth]{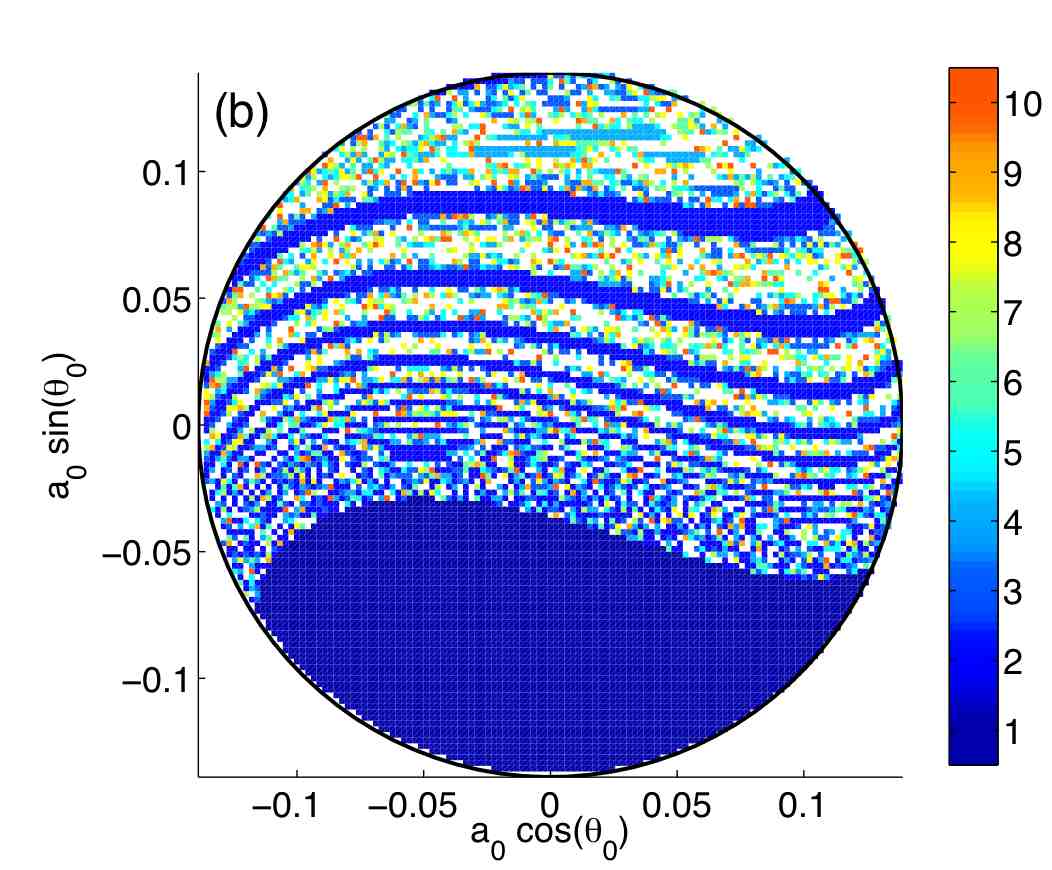}
\caption{(Color online) \subfig{a} The exit velocity and  \subfig{b}  the number of bounces before escape $v_{\rm out}$ as a function of the amplitude and phase of the secondary mode for initial condition~\eqref{eq:morse_init}. Run with parameters $\epsilon=m=1$ and $\omega=2$, and energy $H= m v_0^2/2$, where $v_0=0.287$ is the velocity of the fourth two-bounce window in figure~\ref{fig:ode_in_out}(a).}
\label{fig:odedisk}
\end{center}
\end{figure}

Figure~\ref{fig:odedisk} is the equivalent to figure~\ref{fig:phi4_excited} for a disk of initial conditions of equation~\eqref{eq:morse}.  It is more informative---which we will show later on---to consider a disk of initial conditions of constant energy $H$ given by~\eqref{eq:Hmorse}.   In this figure the point $(x,y)=(A_0 \cos{\theta_0}, A_0 \sin\theta_0)$ corresponds to a numerical simulation to~\eqref{eq:morse} with
\begin{align}
X(0)& =\Xmax; 
&A(0)& = A_0 \cos{(\theta_0-\omega \td)}; \nonumber \\
\dot X(0) &= \sqrt{\frac{2H-\w^2 A^2}{m}};
&\dot A(0) &= \w A_0 \sin{(\theta_0-\omega \td)},
\label{eq:morse_init}
\end{align}
where $\td$ is the time it takes a solution to~\eqref{eq:X_morse} (neglecting $A(t)$) with $(X(0),\dot X(0))$ as given to reach the manifold $\dot X=0$.  The term $\td$ is included so that two adjacent initial conditions on a given radius will arrive at the manifold $\dot X=0$ with approximately the same phase, canceling a very rapid change of this phase as $A_0 \to \sqrt{2 H}/\omega$, the outer edge of the disk.%
\footnote{This set of initial conditions does not lie precisely on a level set of the Hamiltonian, but one on which the $H$ has variations of $O(e^{-\Xmax})$.  This has very little effect on the figure.}

\section{Derivation of the discrete map}
\label{sec:derivation}
Perturbation methods are applicable when $\e\ll 1$. In addition $\w$ may be order-one or else may satisfy $\w\gg 1$. In model~\eqref{eq:morse}, both these conditions are met.  Rigorous analysis in the case that $\e\ll 1$ and $\w=O(1)$ is standard and our results are essentially equivalent to those of Camassa et al. In the case that  $\e \ll 1$ and $\w \gg 1$,  these results may be made rigorous by combining the above work with that of Delshams and Gutierrez~\cite{DelGut:05}.  In either case the formal calculation is the same.  Our method  is an extension of that used by Goodman and Haberman~\cite{GH:04,GooHab:05,GH:05,GooHab:07}.

We let $I = \frac{1}{2}(\dot A^2 + \w^2 A^2)$ be the canonical action variable for the linear oscillator and consider two specific scalings.  First, by setting $\epsilon=0$ in~\eqref{eq:morse}, the two equations decouple completely.  
\begin{subequations}
\label{eq:morse_outer}
\begin{align}
\text{\bf Outer limit:}&  &m \ddot X + U'(X) &=0 \label{eq:X_outer} \\
&&\ddot A + \w^2 A &=0.
\label{eq:A_outer}
\end{align}
\end{subequations}
We will apply this scaling when $X\gg1$.  The second scaling---the inner scaling---is to  let $A=\epsilon A'$.  In the limit $\e\to 0$, this becomes
\begin{subequations}
\label{eq:morse_inner}
\begin{align}
\text{\bf Inner limit:}&  &m \ddot X + U'(X) &=0 \label{eq:X_inner} \\
&&\ddot A' + \w^2 A' + F(X) &=0.
\label{eq:A_inner}
\end{align}
\end{subequations}

The evolution of $X(t)$ is identical in~\eqref{eq:X_outer} and~\eqref{eq:X_inner}.
Its phase portrait contains a homoclinic orbit to a degenerate saddle point at $X=\infty$ which separates bounded, negative energy, solutions from unbounded, positive energy solutions, topologically equivalent to that shown in figure~\ref{fig:AOM}b.  A solution becomes trapped when $\epsilon >0$ and the coupling to the second mode $A(t)$ causes a trajectory to cross from the region of unbounded trajectories to that of bounded, periodic, trajectories.

Equation~\eqref{eq:morse} conserves a time-dependent energy of the form~\eqref{eq:Hmorse}.
Define $E=\frac{m}{2}\dot X^2 + U(X)$,the energy in the mode $X(t)$, so that the level set $E=0$ along the separatrix.  The matched asymptotics will depend on the assumption that the solution remains close to the separatrix throughout its evolution and will alternate between  these two types of approximations.  We define a sequence of times $t_j$, which describes the instants at which the energy jumps between the two modes.  For $t-t_j=O(1)$, the solution is well-approximated by the separatrix orbit, which has the explicit form  $\Xs = \log {\left(1+\frac{(t-t_j)^2}{2m}\right)}$, and that, backwards in time, the oscillatory mode of the approximate solution has the asymptotic behavior 
 \begin{equation}
\label{A_asymptote}
 A(t) \sim \cC\bigl(C_j \cos{\w(t-t_j)} + S_j \sin{\w(t-t_j)}\bigr),
\end{equation}
where $\cC$ is a constant to be defined momentarily.  The immediate objective is to find the asymptotic behavior of $A(t)$ along the inner expansion, and to match this behavior, through the subsequent outer solution, to the behavior along the next inner solution, which is centered at a time $t_{j+1}$, yet to be found.  The linear oscillator can be solved, assuming the solution $X$ can be approximated by the separatrix orbit $\Xs$, using the variation of parameters formula:
\begin{equation}\begin{split}
 A(t) = &\cC\bigl(C_j \cos{\w(t-t_j)} + S_j \sin{\w(t-t_j)}\bigr) + \\
-&\frac{\e}{\w} \sin{\w(t-t_j)} \int_{-\infty}^{t} F(\Xs(\tau-t_j)) \cos{\w(\tau-t_j)} d\tau \\
+&\frac{\e}{\w} \cos{\w(t-t_j)}\int_{-\infty}^{t} F(\Xs(\tau-t_j)) \sin{\w(\tau-t_j)} d\tau, 
\label{eq:A_int}
\end{split}\end{equation}
so that as $t-t_j\to+\infty$,
\begin{equation}\begin{split}
 A(t) \sim &\left(\cC C_j 
 +\frac{\e}{\w}\int_{-\infty}^{\infty} F(\Xs(\tau-t_j)) \sin{\w(\tau-t_j)} d\tau\right)\cos{\w(t-t_j)} \\
 & + \left(\cC S_j -\frac{\e}{\w}  \int_{-\infty}^{\infty} F(\Xs(\tau-t_j)) \cos{\w(\tau-t_j)} d\tau\right)\sin{\w(t-t_j)}.
 \end{split}\end{equation}
As $\Xs(t)$ is an even function, the integral in the coefficient of $\cos{\w(t-t_j)}$ vanishes identically.  Letting
\begin{equation}
\cC  = \frac{\e}{\w}  \int_{-\infty}^{\infty} F(\Xs(\tau)) \cos{\w\tau} \ d\tau
=\frac{\e}{\w}  \int_{-\infty}^{\infty} F(\Xs(\tau)) e^{i\w\tau} d\tau,
\end{equation}
then as $t-t_j \to \infty$,
\begin{equation}
A(t) \sim \cC C_j \cos{\w(t-t_j)}  + \cC (S_j-1)\sin{\w(t-t_j)}.
\label{eq:Aasympt}
\end{equation}
Using the exact form of $\Xs$ and the residue theorem gives
\begin{equation}
\cC  =  \frac{\e}{\w}  \int_{-\infty}^{\infty}\frac{2m}{2m+\tau^2} e^{i\w\tau} d\tau
=   \frac{\e\pi \sqrt{2m}}{\w} e^{-\sqrt{2m}\w}. \label{eq:defC}
\end{equation}
On the ensuing inner approximation, $X(t)\approx\Xs(t-t_{j+1})$, and we write 
$$
 A(t) \sim \cC \left(C_{j+1} \cos{\w(t-t_{j+1})} + S_{j+1} \sin{\w(t-t_{j+1})}\right),
 $$  
 where $t_{j+1}$ remains to be found.  We rewrite equation~\eqref{eq:Aasympt} in the limit as $t-t_{j+1}\to -\infty$ using 
trigonometric identities and find
$
\left(\begin{smallmatrix}C_{j+1}\\S_{j+1}\end{smallmatrix} \right) =
\left(\begin{smallmatrix} \cos\theta_{j+1} & \sin\theta_{j+1} \\ -\sin\theta_{j+1} & \cos\theta_{j+1} \end{smallmatrix} \right)
\left(\begin{smallmatrix}C_j\\S_j-1\end{smallmatrix} \right),
$
where $\theta_{j+1} = \w(t_{j+1}-t_j)$.  Letting $Z_j = C_j + i S_j$, this simplifies to
\begin{equation}
Z_{j+1} = e^{-i\theta_{j+1}} (Z_j-i).
\end{equation}

Before determining $t_{j+1}$, we first examine the evolution of $X(t)$ and its energy $E(t)$.  Along the 
$j$th outer solution (defined as occurring immediately preceding the $j$th inner solution), $X(t)$ and $A(t)$ are uncoupled, and $X_j(t)$ lies along a level surface $E_j = \frac{m}{2} \dot X_j^2(t) + U(X_j(t))$.  Along the the inner solution that follows, we can compute $dE/dt$ and thus $\Delta E$, the change in energy between two successive approaches to the saddle point at infinity.  We calculate
$$
\frac{dE}{dt} = (m\ddot X + U'(X))\dot X = -\epsilon A F'(X) \dot X = -\epsilon A \frac{d}{dt}F(X(t)).
$$
Integrating this along the separatrix yields a Melnikov integral approximation to $\Delta E$, since $F(\Xs(t))\to 0$ as $t\to\pm\infty$,
\begin{equation}E_{j+1} - E_j = \Delta E = -\e \int_{-\infty}^{\infty}  A(t) \frac{d}{dt}F(\Xs(t)) \ dt=  
\e  \int_{-\infty}^{\infty} F(\Xs(t)) \dot A(t) \ dt.
\label{eq:Mel1}
\end{equation}
Using~\eqref{eq:A_int}, this simplifies to
\begin{equation}
\Delta E = \frac{\w^2 \cC ^2}{2}(2 S_j-1).
\label{eq:DE1}
\end{equation}
Defining $E_j = \frac{\w^2\cC ^2}{2} \cE_j$, one finds (even before knowing $\theta_{j+1}$) that the map has a conserved Hamiltonian (derived from~\eqref{eq:Hmorse})
\begin{equation}
\cH = \cE_j + |Z_j|^2.
\label{eq:H}
\end{equation}

An approximation to the interval $(t_{j+1} - t_j)$ can be found using the matching condition for the inner and outer approximations to $X(t)$.  As $t-t_j \to \infty$, the separatrix satisfies
\begin{equation}
t-t_j = \sqrt{\frac m 2} \int_{0}^{X} \frac{dY}{\sqrt{-U(Y)}} = \sqrt{2m} (e^X-1)^{1/2} \sim \sqrt{2m} e^{X/2} + O(e^{-X/2}).
\label{eq:tj}
\end{equation}
Along the near-saddle approach with energy $E_{j+1}$ and letting $t=t^*$ denote the time at which $X$ assumes its maximum value $X^*$ ,  
$$
t^*-t= \sqrt{\frac m 2} \int_{X}^{X^*} \frac{dY}{\sqrt{E_{j+1}-U(Y)}} 
$$
with the asymptotic expansion as $t-t^* \to -\infty$,
\begin{equation}
t^* - t \sim \sqrt{\frac{m}{-2 E_{j+1}} } \cos^{-1} {\left(\sqrt{-2 E_{j+1}} \ e^{X/2} \right)} \sim 
\sqrt{\frac{2m}{-E_{j+1}}}\frac{\pi}{2} - 2\sqrt{m} e^{X/2}.
\label{eq:tstar}
\end{equation}
Summing~\eqref{eq:tj} and~\eqref{eq:tstar},  
$$ t^*-t_{j} \sim \sqrt{\frac{2m}{-E_{j+1}}}\frac{\pi}{2}.$$
A similar calculation yields an identical value for $t_{j+1}-t^*$, so that combining them yields an asymptotic formula for the period
\begin{equation}
\theta_{j+1}= \w T_j = \w(t_{j+1}-t_j) \sim  \w\pi\sqrt{\frac{2m}{-E_{j+1}}}=
\frac{2\pi}{\cC}{\sqrt{\frac{m}{-\cE_{j+1}}}}.
\label{eq:theta}
\end{equation}
Thus, the full map may be written
\begin{equation}\begin{split}
\cE_{j+1} &= \cE_j + 2\Imag Z_j-1\\
Z_{j+1} &= e^{-i\theta_{j+1}}(Z_j-i).
\label{eq:twocomponent}
\end{split}\end{equation}
We may use conservation law~\eqref{eq:H} to eliminate $\cE_j$ from this map
$$
\cE_{j+1}= \cE_j + (2\Imag Z_j-1) = \cH - C_j^2 - S_j^2 +2 S_j -1 = \cH -\abs{Z_j-i}^2,
$$
so the map may be written as a map from the complex plane to itself
$$
Z_{j+1} = e^{\frac{-i \alpha}{\sqrt{|Z_j - i|^2 -\cH}}} (Z_j -i),
$$
where, from equations~\eqref{eq:defC} and~\eqref{eq:theta}, $\alpha = 2\pi\sqrt{m}/\cC =
\sqrt{2}\w e^{\sqrt{2m}\w}/\epsilon$.  Note that as either $\w$ or $\epsilon^{-1}$ will be assumed large, then $\alpha$ will also be large.
Following~\cite{StoKapSir:95}, we change variables to $\cZ_j = Z_j - \frac{i}{2}$ to produce our final form of the map:
\begin{equation}
\label{eq:morsemap}
\cZ_{j+1} \equiv \cF(\cZ_j) = e^{\frac{-i \alpha}{\sqrt{|\cZ_j - i/2|^2 -\cH}}} (\cZ_j -i/2) - i/2.
\end{equation}
Although this makes the map look slightly more complicated, this change of variables has the advantage that the map's inverse is of essentially the same form,
$$
\cZ_{j-1}=\cF^{-1}(\cZ_j) = e^{\frac{i \alpha}{\sqrt{|\cZ_j + i/2|^2 -\cH}}} (\cZ_j +i/2) +i/2,
$$
and simplifies the formulas for fixed points that will follow.  Defining the linear map $\rho(\cZ) =  \cZ^*$, the complex conjugate, then $\cF$ and its inverse are related by:
\begin{equation}
\label{eq:symmetry}
\cF^{-1}=\rho^{-1}\cF\rho.
\end{equation} 

\noindent\textbf{Remark 1} The approach taken here is in a sense orthogonal to the approach for two-dimensional Hamiltonian systems  described by Guckenheimer and Holmes~\cite{GH:83} and applied in our previous studies~\cite{GHW:02,GooHolWei:04}.  Writing the energy as 
$H = H_0(X,\dot X) + I + \epsilon H_1(X,\dot X,\theta)$ where $I$ and $\theta$ are the canonical action-angle coordinates for the $A$-$\dot A$ coordinates, in that approach, one assumes that $\frac{d\theta}{dt}>0$, and uses this fact to reduce the system by replacing the evolution variable $t$ with $\theta$.  One then derives a Poincar\'e map to the section $\theta=\theta_0$.  If, in the uncoupled case $\e=0$, the $X$-$\dot X$ system has a homoclinic orbit, one tries to show using a Melnikov integral that this homoclinic orbit persists in the Poincar\'e map.  One may consider $\cF(\cZ)$ as a Poincar\'e map, where the section is taken to be a subset of the hyperplane $\dot X=0$, switching the roles of $X$ and $A$ from the Guckenheimer and Holmes argument.

\noindent\textbf{Remark 2} The derivation of a similar map for the collective-coordinates model~\eqref{eq:AOM} proceeds in essentially the same manner as for our simplified model~\eqref{eq:morse}.  Evaluation of the (Melnikov) integral~\eqref{eq:defC} defining the scaling constant $\cC$ is significantly harder and cannot be done in closed form, rather as an asymptotic expansion for large $\w$; see~\cite{GooHab:05}.  Additionally, the approximation to the time interval contains an additional $O(1)$ at step~\eqref{eq:theta}: $\theta_{j+1} = C_1/\sqrt{-\cE_{j+1}} + C_2$ for some constants $C_1$ and $C_2$.  The $C_2$-type term is absent from equation~\eqref{eq:theta}.

\section{Interpretation and numerical iteration of the map}
\label{sec:interp}
\subsection{Relation between map and ODE}
\label{sec:map_ode}
The map $\cF$ defined by~\eqref{eq:morsemap} greatly compresses the information content of the ODE system~\eqref{eq:morse}, and it is worth  discussing the correspondence between the dynamics on the $\cZ$-plane and that of the full ODE.  Map~\eqref{eq:morsemap} depends on two constant parameters $\alpha$ and $\cH$.   The total change of phase of $a(t)$ between two consecutive interactions is directly proportional to $\alpha$, and thus to $\w$.  The constant $\cH$ is a rescaling of the Hamiltonian $H$. 
Solutions on level sets $\cH <0$ may never escape, as the energy~\eqref{eq:Hmorse} is positive-definite.  Solutions on the level set $\cH = 0$ may escape to $\infty$, but only along the separatrix orbit,with escape velocity approaching zero. Almost every solution with $\cH>0$  escapes to $\infty$ at finite velocity as $t\to\pm\infty$.

Map~\eqref{eq:morsemap} is extremely similar to the Ikeda map, 
$$
W_{j+1} = {\mathcal G}(W_j) \equiv b e^{i/( |W_j|^2 + 1)} + \epsilon.
$$
which arises in the modeling of lasers.  Here $0<b\le1$ is a dissipation factor, $W_j\in \mathbb{C}$ and $\epsilon \in \mathbb{R}$.  This is the composition of three operations: nonuniform rotation about the origin, contraction by a factor $b$, and translation by $\epsilon$.  If $b=1$, $\mathcal G$  is the composition of two area- and orientation-preserving maps and thus preserves both area and orientation.  

Stolovitzky et al.\ studied a similar map of the form 
$$
W_{j+1} = {\mathcal G}(W_j) \equiv b e^{i/|W_j|^2} + \epsilon
$$
which, like the map $\cF$, is singular~\cite{StoKapSir:95}.  Map $\cF$ is also formed as the composition of uniform translation and nonuniform rotation---whether rotation takes place first, last, or in the middle, is equivalent dynamically.

When $\cH<0$, the map $\cF$ is well-defined, continuous, and invertible on all of $\mathbb{C}$.  At $\cH = 0$, the rotation rate diverges at $\cZ = i/2$.
When $\cH >0$, the domain of map~\eqref{eq:morsemap} is  the complement of the disk 
$$
\cDo = \left\{\cZ : \abs{\cZ-\frac{i}{2}}^2 \le \cH \right\}.
$$
$\cF$ is continuous outside $\cDo$, but the rotation rate diverges as the disk is approached.  For points in $\cDo$, the two-component form of the map, system~\eqref{eq:twocomponent}, the next iterate $\cE_{j+1}$ is well-defined and positive, but the angle $\theta$ and thus likewise $C_{j+1}$ and $S_{j+1}$ are undefined.

\begin{figure}
\begin{center}
\includegraphics[width=3in]{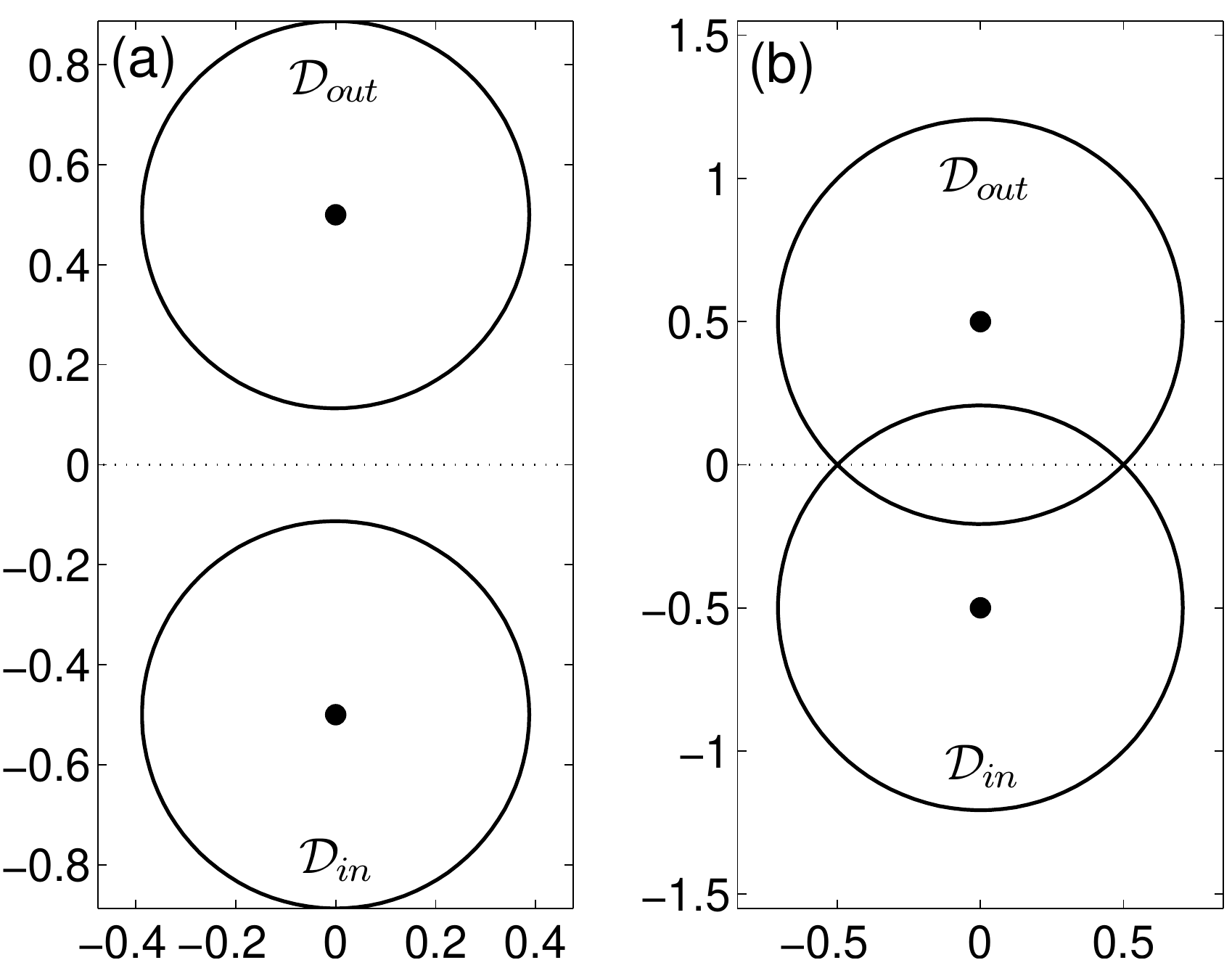}
\caption{The disks $\cDi$ and $\cDo$.  When $0<\cH<\frac{1}{4}$, the disks do not intersect. When $\cH>\frac{1}{4}$, the disks intersect. In \subfig{a} $\cH = 0.15$, and in \subfig{b}, $\cH = 0.5$.}
\label{fig:D_in_out}
\end{center}
\end{figure}

The map's range is the complement of the disk
$$
\cDi = \left\{\cZ : \abs{\cZ+\frac{i}{2}}^2 \le \cH \right\}=\rho \cDo;
$$
see figure~\ref{fig:D_in_out}.    Note, from conservation law~\eqref{eq:H}, when $\cE > 0$, $\abs{\cZ+i/2}^2 < \cH$, meaning that points inside the disk $\cDi$ correspond to portions of the trajectory \emph{outside} the separatrix of figure~\ref{fig:AOM}b in differential equations~\eqref{eq:X_morse}.  Similarly, when $\cZ$ is outside the disk $\cDi$, the trajectory of~\eqref{eq:X_morse} are inside the separatrix.  Note in figure~\ref{fig:odedisk}b, the set of initial conditions that escape on the first iterate ($n_{\rm bounces}=1$) is bounded by two arcs, the boundary of $\cDi$ and another that is nearly a circular arc on the interior of $\cDi$.  As $\epsilon\to 0^+$ , the approximation made in deriving $\cF$ are become more accurate, and this second boundary becomes more circular.

Of primary interest is the evolution of points in $\cDi$.  These points that correspond to solitons starting at a distance $\infty$ from the collision point.  The map is iterated until the trajectory lands in the circle  $\cDo$ and the soliton escapes.  How many iterates this takes then tells how many bounces the solitary wave undergoes before escaping.   The initial condition $\cZ_1=-i/2$ (at the center of $\cDi$) describes a trajectory along which $A(t) \to 0$ as  $t\to -\infty$.  If, for some $n>0$, $\cZ_n = i/2$ (at the center of $\cDo$) , then the solution escapes to $X=\infty$ after $n$ in such manner that $A(t) \to 0$ as $t\to +\infty$.  Thus, the energy level $\cH$ contains an $n$-bounce resonance if $\cF^{n-1}(-i/2)=i/2$.  Also of interest are the iterates  initial conditions $\cZ \in \partial \cDi$.  These correspond to trajectories along which $(X,\dot X) \to (\infty,0)$ as $t \to -\infty$.  Similarly if $\cZ_n \in  \partial \cDo$, the trajectory will approach $(X,\dot X) = (\infty,0)$ as $t \to +\infty$.  The existence of a point  $\cZ_1 \in \partial \cDi$ such that $\cF^{n-1} \in \partial \cDo$, indicates that there exists an orbit homoclinic to a periodic orbit with energy $\cH$.

On the energy levels $\sqrt{\cH} > 1/2$, the disks $\cDi$ and $\cDo$ have an intersection of nonzero measure.  Points in the intersection correspond to solitary waves of sufficient energy never to be trapped (they come from $X = \infty$ and return to $X = \infty$ with only one application of the map).  The two points on $\partial \cDi \cup \partial \cDo$ correspond to one-pulse homoclinic orbits to some periodic orbit. 
The energy level $\cH=1$ is the energy level of the critical velocity for capture of an unexcited wave.  At any higher energy the point $\cZ=-i/2$ which corresponds, recall, to the solitary wave arriving from $X=\infty$ with no energy in the internal mode is in $\cDo$, and the soliton will escape without ever being captured.

\subsection{Numerical iteration}
\label{sec:iteration}

\begin{figure}
\begin{center}
\includegraphics[width=0.5\textwidth]{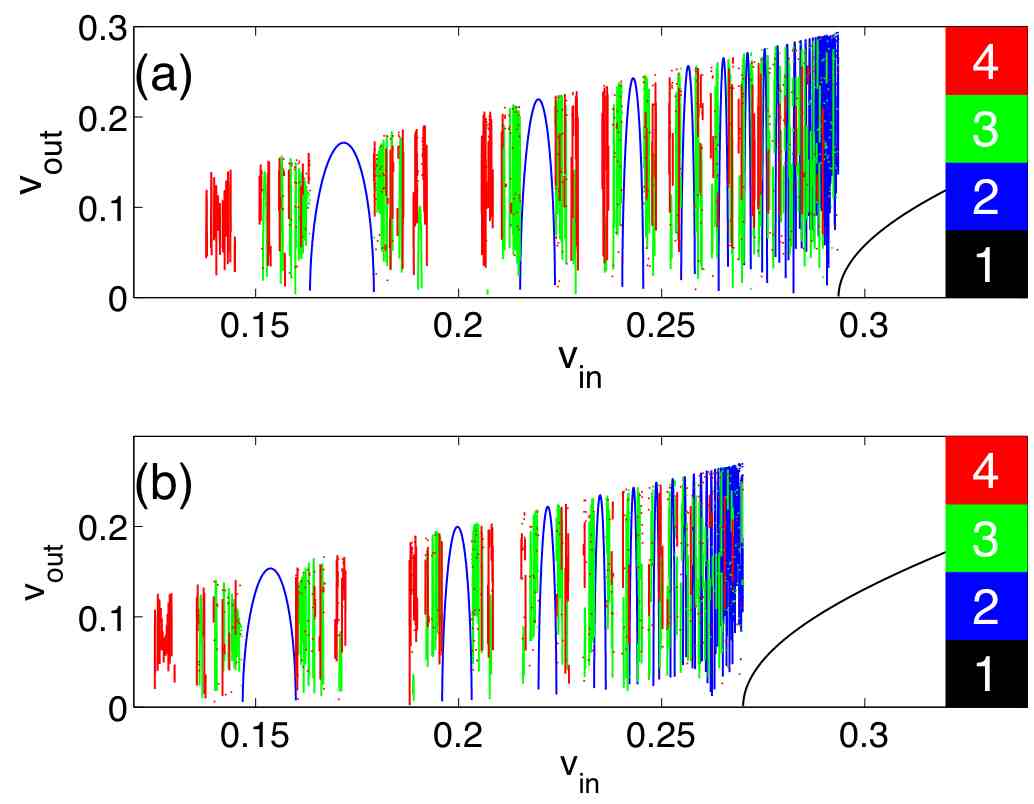}
\caption{(Color online) The outgoing speed as a function of the incoming speed for \subfig{a} the ODE~\eqref{eq:morse} and \subfig{b} the map~\eqref{eq:morsemap} suitably scaled.}
\label{fig:inout_compare}
\end{center}
\end{figure}

We display numerical iterations of map $\cF$. Figure~\ref{fig:inout_compare} shows that map~\eqref{eq:morsemap} can reproduce the figure~\ref{fig:ode_in_out}a, with parameters $\epsilon=0.25$, $m=1$, $\omega=1$.  Each point on the $x$-axis in figure~\ref{fig:inout_compare}(b) corresponds to the initial condition $(\cE_0,Z_0)= m v_{\rm in}^2 / \w^2\cD^2/,0)$ in the map's two-component form~\eqref{eq:twocomponent}, while $v_{\rm out}\propto\sqrt{\cE_n}$ where the solution escapes to infinity on the $n$ iteration. For these parameter, the critical velocity is computed correctly to about 9\%, and the map calculation reproduces well the topological structure of the ODE simulations.  Quantitative agreement between the map and the ODE simulations can be improved by using the value of $E_{j+1}-E_j = \frac{m \vc^2}{2}$, where $\vc$ is obtained from direct numerical simulation rather than from the Melnikov integral computation~\eqref{eq:Mel1}.

\begin{figure}
\begin{center}
\includegraphics[width=\textwidth]{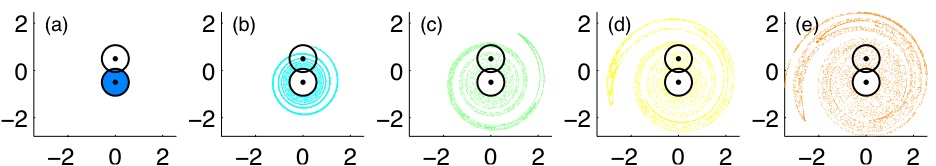}
\caption{(Color online) The initial conditions in the disk $\cDi$, subfigure \subfig{a} and their first four iterates under map~\eqref{eq:morsemap}, subfigures \textbf{(b)-(e)}.}
\label{fig:iterates}
\end{center}
\end{figure}

\begin{figure}
\begin{center}
\includegraphics[height=0.4\textwidth]{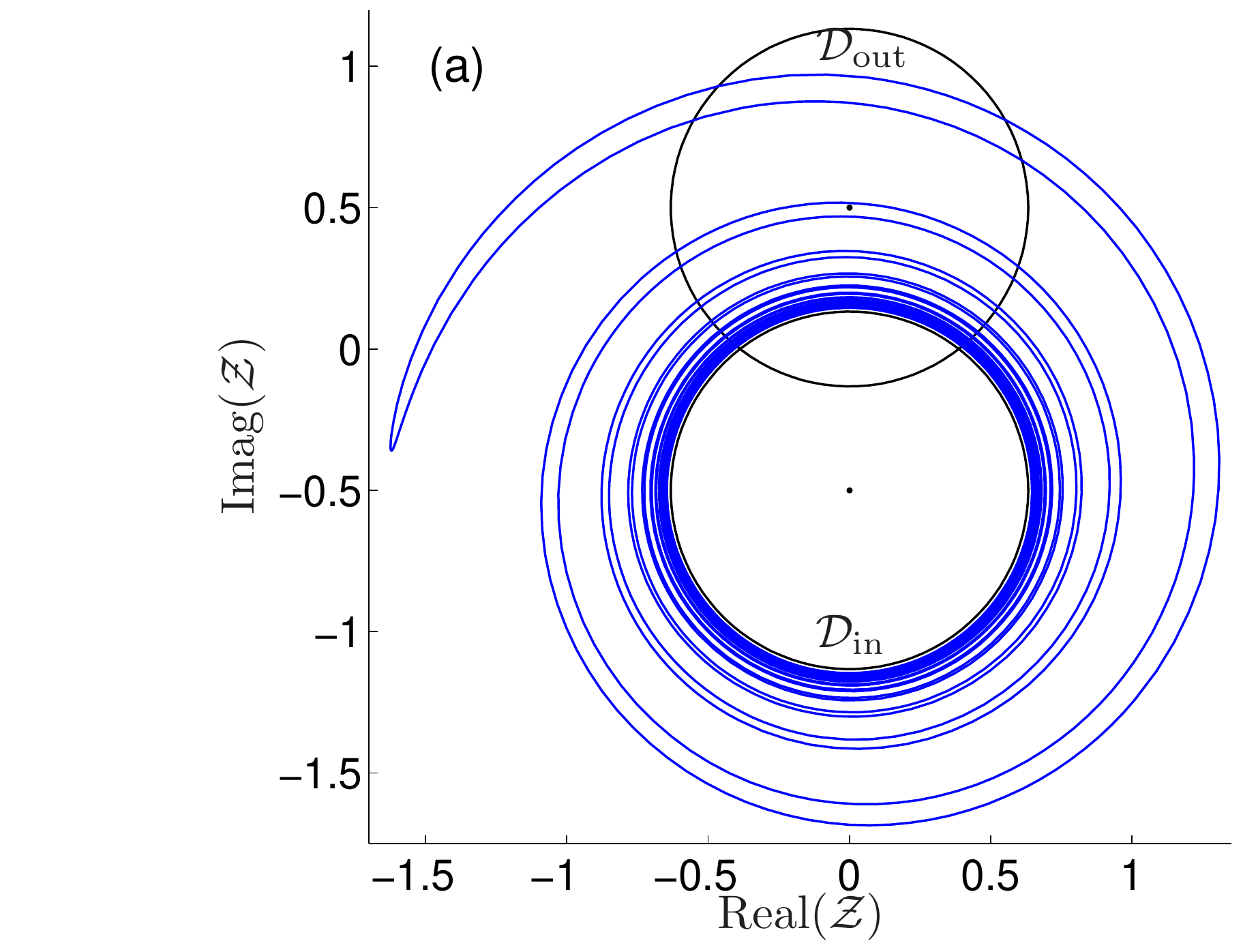}
\includegraphics[height=0.4\textwidth]{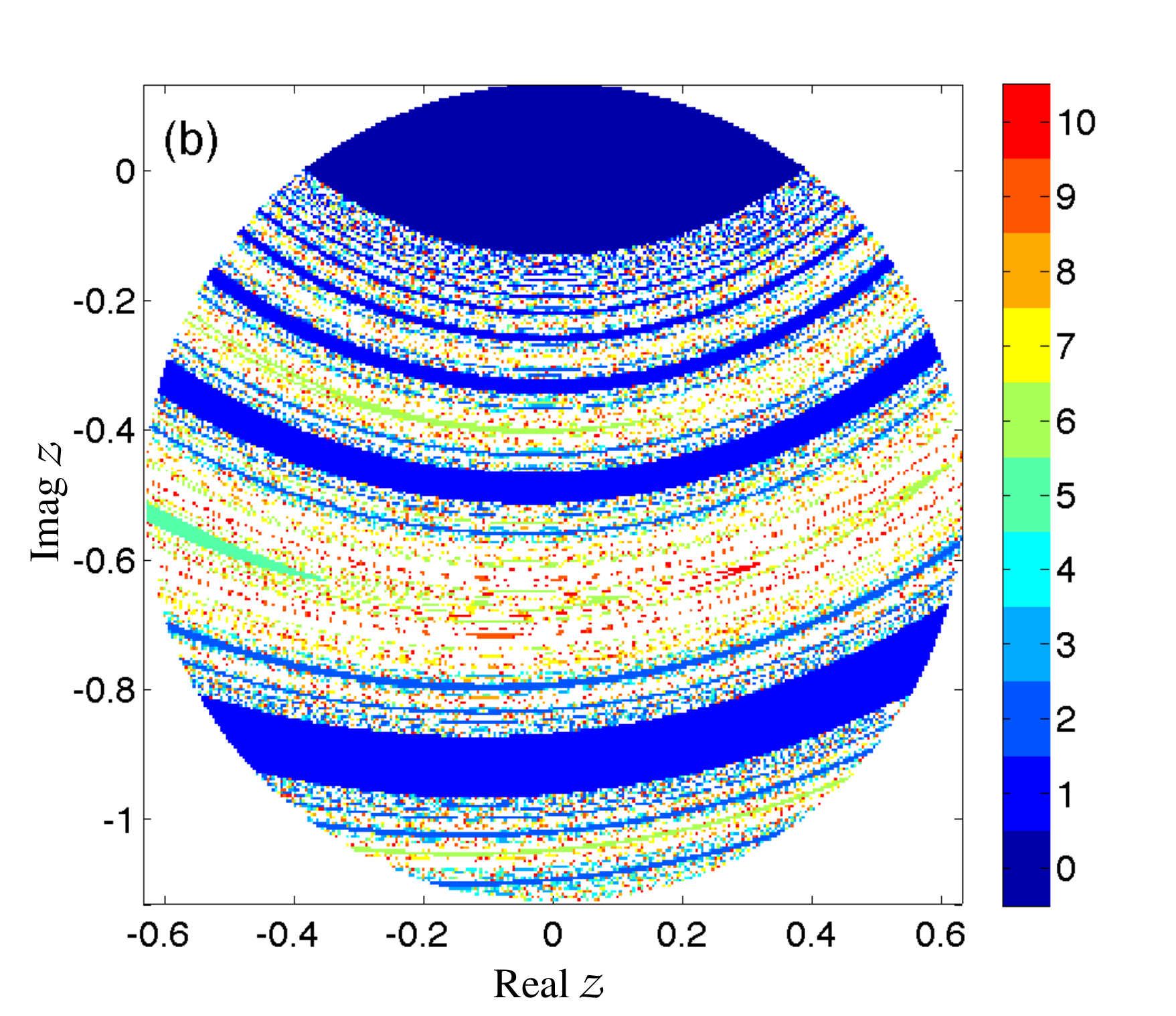}
\caption{(Color online)  Left: The curve $\cF(\partial\cDi)$. Right: The number of iterates preceding escape for points in the disk $\cDi$. }
\label{fig:edge}
\end{center}
\end{figure}

Figure~\ref{fig:iterates} shows the first four iterates of the disk $\cDi$ under map~\eqref{eq:morsemap}, with the same parameter values as figure~\ref{fig:inout_compare} and energy level corresponding to the initial velocity marked by~$\star$ in that figure.  This gives the parameter values $\alpha=23.27$ and $\cH=0.324$ in~\eqref{eq:morsemap}. Figure~\ref{fig:edge}(a) shows a portion of the curve $\cF(\partial \cDi)$ which wraps around $\cDi$ infinitely many times. This exhibits the stretching and folding typical of chaotic systems in addition to the singular behavior near near the outside edge of $\cDi$.  Fgure~\ref{fig:edge}b, shows the number of iterates preceding escape depends sensitively on the initial condition in the disk $\cDi$, showing qualitative agreement with figure~\ref{fig:odedisk}.x

\section{Analysis of the iterated map}
\label{sec:morse}

\subsection{Analogy to Sitnikov's reduced 3-body problem}
\label{sec:sitnikov}
Moser considers the reduced 3-body problem, due originally to Sitnikov~\cite{Mos:73}. In this problem two primaries of identical mass $M$ orbit about their joint center of mass in planar elliptical orbits of eccentricity $\epsilon$.  A third body moves in a line normal to this plane and through the center of mass of the two primaries and evolves under attraction to the two primaries.  The third body is assumed to exert negligible force on the first two.  The third mass satisfies the differential equation 
\begin{equation}
\label{eq:sitnikov}
\ddot z = -\frac{z}{(z^2+\rho^2)^{3/2}},
\end{equation}
where $\rho = \frac{1}{2}-\epsilon \cos{t} + O(\epsilon^2)$.

He then considers the sequence of times $t_n$ at which $z=0$ and derives a map
\begin{equation}
\label{eq:mosermap}
(t_{n+1},v_{n+1}) = \cG_\epsilon(t_n,v_n),
\end{equation}
where $v_n = \abs{\dot z(t_n)}$.   In the limit $\epsilon \to 0^+$, the differential equation~\eqref{eq:sitnikov}, and thus of the map $\cG_0$, as well, is completely integrable.  In fact, the map $\cG_0$ reduces to 
$$ v_{n+1} = v_{n} = v_0; \quad t_{n+1}=t_n+T(v_n) = t_0 + (n+1)T(v_0),$$
where the time delay $T(v)$ is an increasing function of $v$ such that $T(0)=0$, $T(v)\to\infty $ as $v \to 2^-$ and $T(v)$ is undefined for $v\ge2$. Letting the variables $t$ and $v$ be polar coordinates for $\mathbb{R}^2$,  $\cG_0$ maps the circle of radius two to itself, with a twist angle that diverges as the circumference is approached.  The map is undefined for $v>2$, which correspond to orbits that travel monotonically from $z=\pm \infty$ to $z=\mp \infty$, and for which the notion of time between successive zeros is meaningless.  When $0<\epsilon\ll 1$, the domain and range of $\cG_\epsilon$ no longer coincide.  The map $\cG_\epsilon$ is defined on an ellipsoidal region $D_0$ and maps onto a second ellipse $D_1=\cG_\epsilon D_0$.  These regions are close analogs to the complements of $\cDo$ and $\cDi$. As $\epsilon \to 0^+$, both regions approach the circle of radius two centered at the origin.

\begin{figure}
\begin{center}
\includegraphics[width=2in]{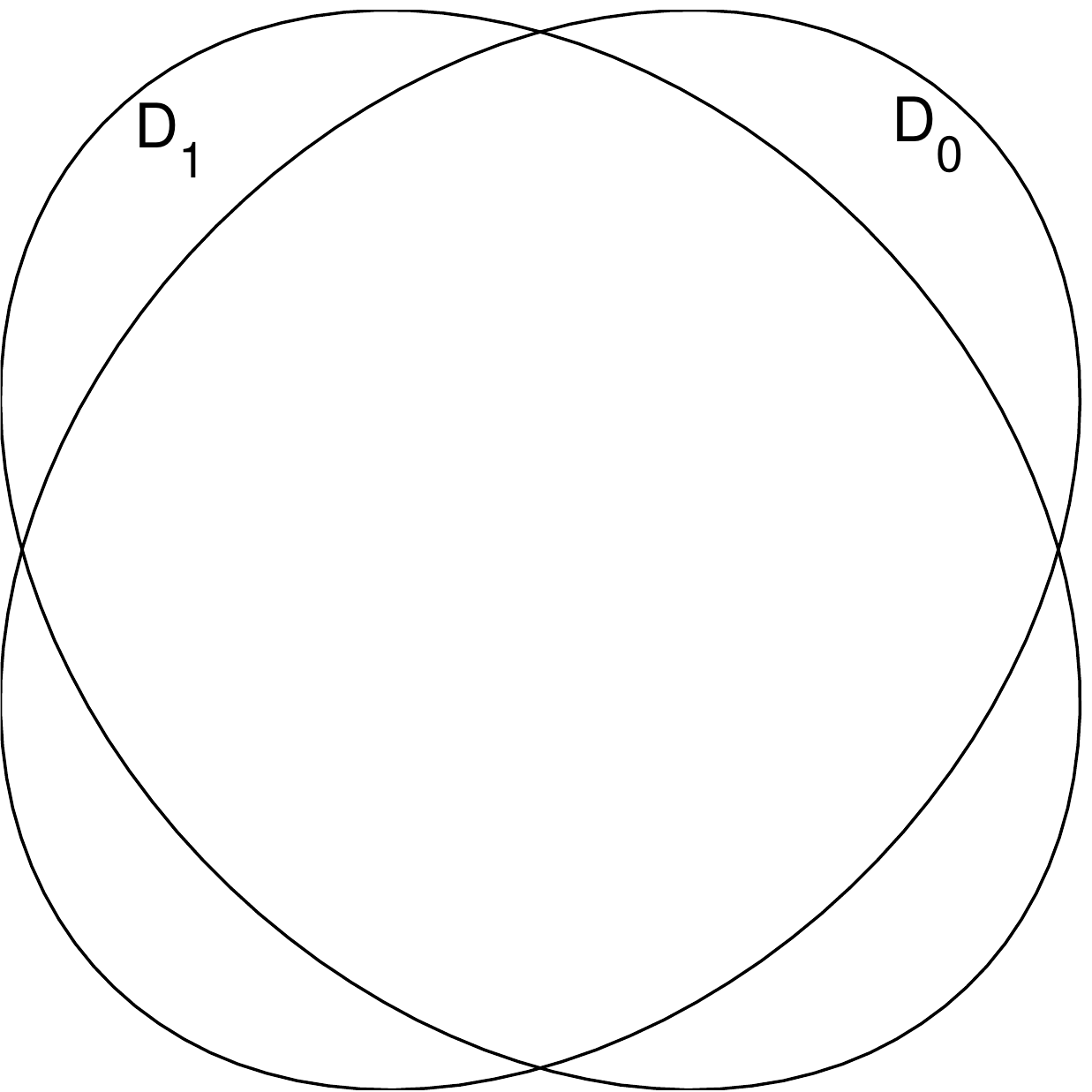}
\caption{The domain $D_0$ and range $D_1$ of map~\eqref{eq:mosermap}.}
\label{fig;moser}
\end{center}
\end{figure}

\begin{figure}
\begin{center}
\includegraphics[width=2in]{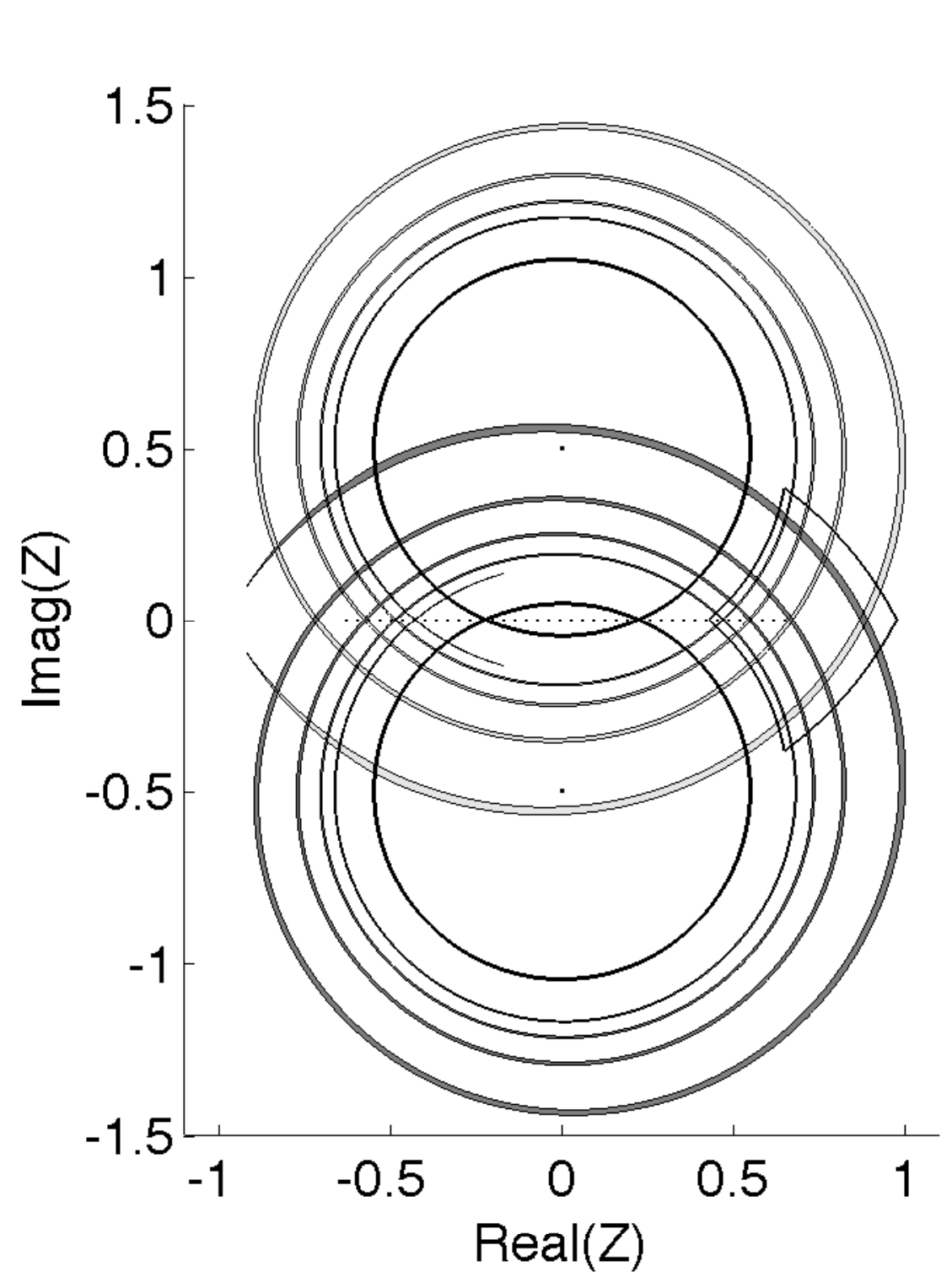}
\includegraphics[width=2in]{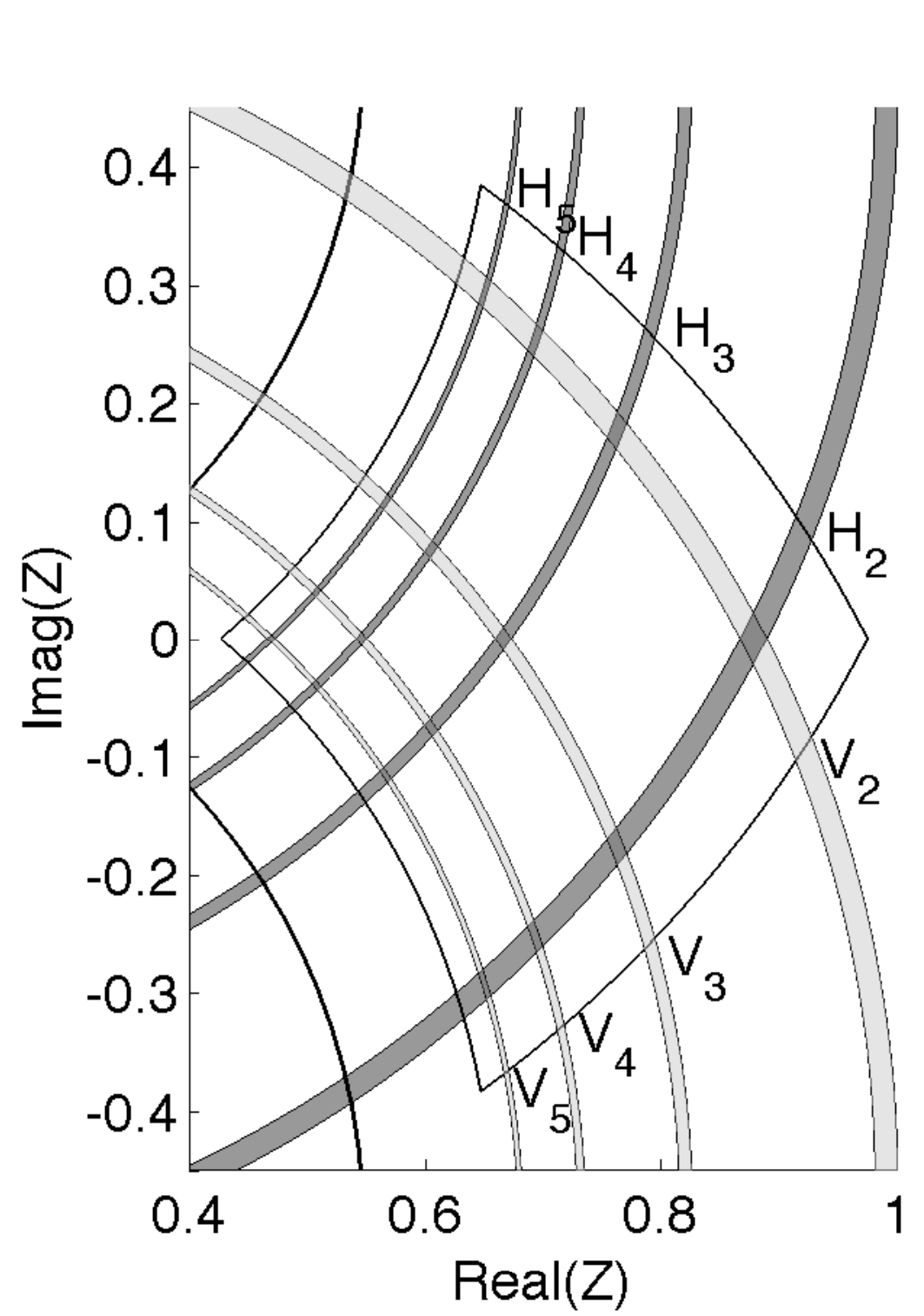}
\caption{The construction of the horseshoe, with the set $\cF(\cR)$ shown in light gray and the set $\cF^{-1}(\cR)$ in dark gray. \subfig{a} full image; \subfig{b} closeup on the rectangle $\cR$.}
\label{fig:horseshoe}
\end{center}
\end{figure}

Moser~\cite{Mos:73} proves the following theorem, after defining the integers 
$$s_k=\left\lfloor \frac{t_{k+1}-t_k}{2\pi} \right\rfloor$$
as the number of complete revolutions made by the two primaries between two consecutive zeroes of $z(t)$.
\begin{theorem}
\label{thm:moser}
Given a sufficiently small eccentricity $\epsilon>0$, there exists an integer $m=m(\epsilon)$ such that any sequence $s$ with $s_k>m$ corresponds to a solution to the differential equation~\eqref{eq:sitnikov}.
\end{theorem}
Notice that the sequence $s_k$ can be chosen completely arbitrarily. Further, one may define semi-infinite or finite sequences that begin or end with $\infty$, corresponding to solutions that arrive from or escape to $z=\pm \infty$.  This theorem is proven by constructing a horseshoe on which one defines a Bernoulli shift on a countably infinite number of symbols. In figure~\ref{fig:horseshoe}, we construct such a horseshoe for map~\eqref{eq:morsemap} by considering the image of the topological rectangle
$$
\cR= \left \{\cZ: \sqrt{\cH}<r_1 < \abs{\cZ+\frac{i}{2}} < r_2 \right\} \cap
\left \{\cZ: \sqrt{\cH}<r_1 < \abs{\cZ-\frac{i}{2}} < r_2 \right\}\cap
\left \{ \Real{\cZ}>0 \right\}.
$$
Here we have set $\alpha=15$, $\cH=0.3$, $r_1 = 1.2\sqrt{\cH}$, $r_2 = 2\sqrt{\cH}$. The set $\cF(\cR))$ ($\cF^{-1}(\cR)$, respectively) wraps about four times around the disk $\cDo$ ($\cDi$, respectively). $\cF(\cR)$ intersects $\cR$ in four stripes labeled $V_2$ through $V_5$ ($H_2$ through $H_5$ respectively).  The subscript is given by $\lfloor \alpha/\sqrt{\abs{Z-i/2}^2 -\cH}\rfloor$ for a point on the intersection of the stripe and the real $\cZ$ axis and represents the number of complete oscillations made by $A(t)$ between two consecutive bounces.  

The existence of this horseshoe implies the existence of a Cantor set of points which remain in the rectangle $\cR$ for all iterations of $\cF$ or $\cF^{-1}$ on which the action of the map is equivalent to a Bernoulli shift on four symbols.%
\footnote{In fact to show the existence of chaos is sufficient to take the rectangle $\cR$  smaller, so that it only contains the four stripes $V_2$, $V_3$, $H_2$ and $H_3$.  One may show the existence of a horseshoe without explicitly constructing it using Melnikov integral arguments alone, as in~\cite{DanHol:95}.}
As $r_1 \nearrow \sqrt{H}$, the number of stripes contained in $\cR$ grows without bound, so in this limit, the Bernoulli shift operator requires a countably infinite number of symbols.  This is what allows one to specify the sequence of integers describing the number of complete oscillations between consecutive zeros in Theorem~\ref{thm:moser}.  Such ``infinite horseshoes'' are the subject of a recent preprint by Zambrano et al~\cite{ZamSanKen:07}.

We may find an integrable limit of the map~\eqref{eq:morsemap} by defining $\delta = \frac{1}{2\sqrt{\cH}}$, $\gamma=\frac{\alpha}{\sqrt{\cH}}$ and $\cY = \frac{\cZ}{\sqrt{\cH}}$, in which case 
$$
\cY_{n+1} = e^{-i \frac{\gamma}{\sqrt{\abs{\cY_n-i\delta}^2-1}}}\cY_n-i\delta.
$$
Then as $\delta \to 0^+$ with $\gamma>0$ fixed, map~\eqref{eq:morsemap} approaches the integrable iteration
$$
\cY_{n+1} = e^{-i\frac{\gamma}{\sqrt{\abs{\cY_n}^2-1}}}\cY_n.
$$
In this limit $\cDo$ and $\cDi$ degenerate to the unit circle.  Letting $\cX= \cY^{-1}$, this map is identical to $\cG_0$.

\subsection{$n$-bounce resonant solutions and multipulse homoclinic orbits}
\label{sec:nbounce}
The analysis of section~\ref{sec:derivation} can be used to determine some key features of the dynamics.  Some of these results appear in our earlier work, but have a nice geometrical interpretation in the present context.
First,  the critical velocity for capture is found using~\eqref{eq:DE1}.  In the case that the internal mode is initially unexcited ($\cZ_0=-i/2$), one finds that $\Delta E = - \frac{\w^2 \cC^2}{2}$, so that the critical velocity for capture is 
$$\vc = \w \cC/\sqrt{m} =   \e\pi \sqrt{2} e^{-\sqrt{2m}\w}.$$  We define an $n$-bounce resonant solution  by the property $\cZ_0=-i/2$ and $\cZ_{n-1}=i/2$, so that the map is undefined and the solitary wave escapes on the $n$ iteration and no energy remains in the oscillating mode as $t\to\pm\infty$.%
\footnote{Note that the map as defined one component form~\eqref{eq:morsemap} is undefined for $\cZ_{n-1}=i/2$, but that in the two component form---equations~\eqref{eq:theta} and~\eqref{eq:twocomponent}---this gives $\cE_n=\cE_0$.}  %
This may conveniently be written, using formula~\eqref{eq:symmetry}, as
\begin{equation}
\label{eq:nbounce}
\cF^{n-1}\left(\frac{-i}{2}\right) = \frac{i}{2}.
\end{equation}
In the case $n=2m$, 
\begin{equation}
\cF^{m}\left(\frac{-i}{2}\right) = \cF^{-(m-1)}\left(\frac{i}{2}\right) =\rho\cF^{m-1}\left(\frac{-i}{2}\right)
\label{eq:evenbounce}
\end{equation}
and in the case $n=2m+1$
\begin{equation}
\cF^{m}\left(\frac{-i}{2}\right) = \cF^{-m}\left(\frac{i}{2}\right)=\rho\cF^{m}\left(\frac{-i}{2}\right).
\label{eq:oddbounce}
\end{equation}
The two sides of equation~\eqref{eq:oddbounce} are complex conjugates of each other, so, equivalently 
$$
\cF^m\left(\frac{-i}{2}\right) \in \R.
$$
We illustrate this with explicit formulae for the 2- and 3-bounce initial velocities, and an implicit formula which yields the 4-bounce initial velocities.  We first define the quantities (assuming the quantities under any square roots signs are positive)
$$\phi = \alpha/\sqrt{1-\cH},$$
$$\cZ_1=\cF\left(\frac{-i}{2}\right) =-i e^{-i\phi}-\frac{i}{2},$$ and
\begin{equation}
\label{eq:psi}
\psi= \alpha/\sqrt{4 \cos^2{\frac{\phi}{2}}-\cH}= \alpha/\sqrt{4 \cos^2{\frac{\alpha}{2\sqrt{1-\cH}}}-\cH}.
\end{equation}
Then the condition for a two-bounce resonant solution, from~\eqref{eq:evenbounce} is
$\cZ_1 = \frac{i}{2}$, for a three bounce resonant solution is $\cF(\cZ_1)\in \R$, and for four-bounce resonant solution is $\cF(\cZ_1)=\cZ_1^*$.

Solving for the two-bounce resonance leads to the algebraic condition 
\begin{equation}
\label{eq:2b}
\frac{\alpha}{\sqrt{1-\cH}}= (2n-1)\pi;
\end{equation}
for the three-bounce resonance,  
\begin{equation}
\label{eq:3b}
\frac{\alpha}{\sqrt{1-\cH}}= \left(2n-1\pm \frac{1}{3}\right)\pi;
\end{equation}
and for the four-bounce resonance we find
\begin{equation}
\label{eq:4b}
\phi + \psi = (2n-1)\pi.
\end{equation}

Solving equations~\eqref{eq:2b} and~\eqref{eq:3b} for $\cH$ with fixed $\alpha$ allows us to find the two-bounce resonance velocities
\begin{equation}
\label{eq:2bounce}
v_{2,n}=\sqrt{\vc^2-\frac{4\w^2}{(2n-1)^2}}
\end{equation}
and three-bounce resonance velocities
\begin{equation}
\label{eq:3bounce}
v_{3,n\pm}=\sqrt{\vc^2-\frac{4\w^2}{(2n-1\pm \frac{1}{3})^2}}.
\end{equation}
One cannot derive a closed-form expression for the four-bounce resonance velocities.

\begin{figure}
\begin{center}
\includegraphics[width=3in]{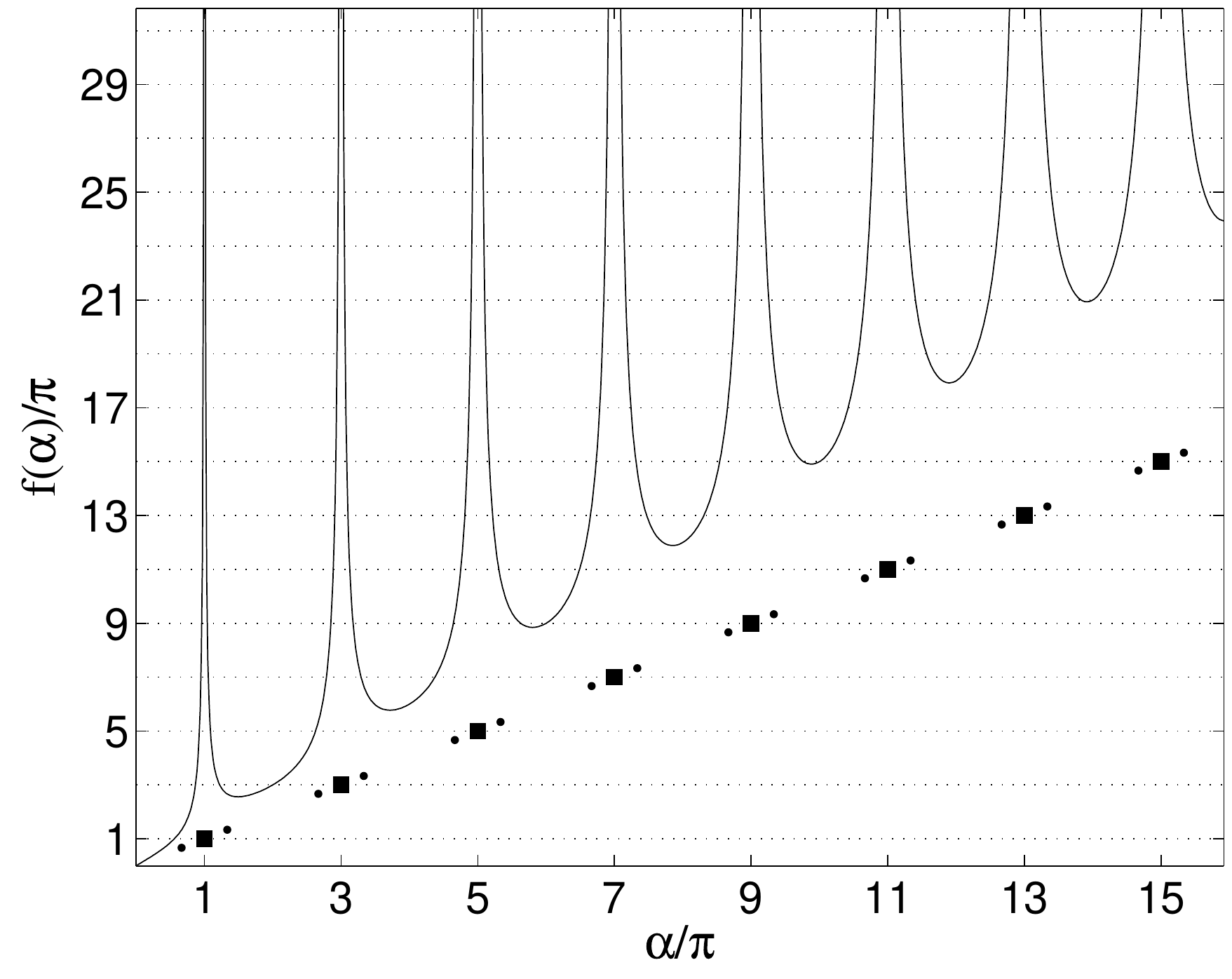}
\caption{The locations of two-, three-, and four-pulse homoclinic orbits, (squares, dots, and intersections between the solid curve and the dotted horizontal lines).}
\label{fig:homoclinic_sequence}
\end{center}
\end{figure}

Solutions to equation~\eqref{eq:nbounce} on the energy level $\cH=0$ correspond to multi-pulse heteroclinic orbits of ODE system~\eqref{eq:morse}.%
\footnote{To be more precise if such a solution exists, then, by the implicit function theorem, there exists a multipulse homoclinic orbit for a nearby value of $\alpha$.}
Solving~\eqref{eq:2b} for $\alpha$ when $\cH=0$ gives a condition for the existence of a two-bounce homoclinic orbit to $\infty$:
\begin{align}
\label{eq:2bounce_hom}
\alpha&= \sqrt{2}\w e^{\sqrt{2}m\w}=(2n-1)\pi
\intertext{and, using~\eqref{eq:3b}, the three-bounce homoclinic orbits satisfy}
\label{eq:3bounce_hom}
\alpha&= \sqrt{2}\w e^{\sqrt{2}m\w}=\left(2n-1\pm\frac{1}{3}\right)\pi.
\end{align}
Most interestingly, equation~\eqref{eq:4b} becomes
$$
\alpha \left( 1 + \frac{1}{2\abs{\cos{\frac{\alpha}{2}}}} \right) = (2n-1)\pi.
$$
The term $ \frac{1}{2\abs{\cos{\frac{\alpha}{2}}}}$ diverges as $\alpha \to (2m-1)\pi$, which means that this equation has an infinite number of solutions as $\alpha \to (2m-1)\pi$ from above or below.  Thus, between any pair of two-bounce homoclinic orbits there exists a pair of three-bounce homoclinic orbits and a countably infinite sequence of four-bounce homoclinic orbits.  This is summarized in figure~\ref{fig:homoclinic_sequence}.  A similar pattern holds for two-, three-, and four-bounce resonances.

\subsection{Fixed points and periodic orbits}
\label{sec:fixpts}
Fixed points $\cZ_f$ of~\eqref{eq:morsemap} must satisfy 
$$\abs{\cZ_f+\frac{i}{2}} = \abs{\cZ_f -\frac{i}{2}},$$
i.e.\  $\cZ_f \in \R$.  Using trigonometric identities and setting $\cZ_f = C_l$, we find the fixed points satisfy
$$
C_l = -\frac{1}{2}\cot{\left (\frac{\theta_l}{2}-\pi l\right)}
$$
where $\theta_l = \alpha/\sqrt{  C_l^2 + \frac{1}{4} - \cH}$, and we assume $2\pi l<\theta_l<2\pi(l+1)$.  This has at most one root for each value of $l$, and allows for a convenient indexing of the branches of fixed points $C_l(\alpha,\cH)$.   A more useful form for calculations is
\begin{equation}
\label{eq:fixpt}
2 C_l \sin{\frac{\theta_l}{2}} + \cos{\frac{\theta_l}{2}} = 0.
\end{equation} 

The stability of a fixed point is determined by the eigenvalues of $\cJ(\cZ_f)$, the Jacobian matrix.%
\footnote{As map~\eqref{eq:morsemap} is not an analytic function of $\cZ$.  The Jacobian is taken with respect to the real and imaginary parts of $\cZ$.}
  Map~\eqref{eq:morsemap} is orientation- and area-preserving,  thus its Jacobian has unit determinant.  Thus the stability is determined entirely by $\tau=\trace{\cJ(\cZ_f)}$.  If $\abs{\tau}<2$, the fixed point is (neutrally) stable; if not, it is unstable.  If $\tau>2$, then $\cZ_f$ is a saddle point, and if $\tau<-2$, $\cZ_f$ is a saddle with reflection (both eigenvalues negative).  At  points where $\tau=2$ one may find the canonical bifurcations, in this case saddle-nodes, and points where $\tau=-2$ correspond to period-doubling bifurcations.  We may find most of these bifurcation values directly.
  
  The trace is given by
\begin{equation}
\label{eq:tau}
\tau=  \frac{\alpha \left(C_l^2+\frac{1}{4}\right)}{\left(C_l^2+\frac{1}{4}- \cH\right)^{3/2}} \sin{\theta_l} + 2\cos{\theta_l}.
\end{equation}
At a saddle-node bifurcation, $\tau=2$, which simplifies~\eqref{eq:tau} to
\begin{equation}
\sin{\frac{\theta_l}{2}}\left( \frac{\alpha R}{2 \left(R-\cH\right)^{3/2}} \cos{\frac{\theta_l}{2}} 
- \sin{\frac{\theta_l}{2}}\right)=0.
\label{eq:tau_saddlenode}
\end{equation}
The solution $\sin{\frac{\theta_l}{2}}=0$ is inconsistent with~\eqref{eq:fixpt}.  Setting the other factor zero is equivalent to 
\begin{equation}
\label{eq:saddlenode}
\cH = C_l^2 + \frac{1}{4}- \left(\alpha \left(C_l^2 + \frac{1}{4}\right) C_l\right)^{2/3}.
\end{equation}
At a flip (period-doubling) bifurcation, $\tau = -2$, which simplifies~\eqref{eq:tau} to
\begin{equation}
\cos{\frac{\theta_l}{2}}\left( \frac{\alpha R}{2 \left(R-\cH\right)^{3/2}} \sin{\frac{\theta_l}{2}} 
+ \cos{\frac{\theta_l}{2}}\right)=0.
\label{eq:tau_flip}
\end{equation}
This has solutions of two types.  If $\cos{\frac{\theta_l}{2}}=0$, then $\theta=(2l+1)\pi$ and equation~\eqref{eq:fixpt} further implies that $C_l=0$.  The locations of these bifurcations are then described by
\begin{equation}
\label{eq:flip1}
\cH = \frac{1}{4} - \left(\frac{\alpha}{(2l+1)\pi}\right)^2.
\end{equation}
Flip bifurcations can also happen where $\cos{\frac{\theta_l}{2}}\neq0$ but the other term in~\eqref{eq:tau_flip} vanishes. This can be shown to imply
\begin{equation}
\label{eq:flip2}
\cH=  C_l^2 + \frac{1}{4} -\left( \frac{(C_l^2+\frac{1}{4})\alpha}{4C_l}\right)^{2/3}.
\end{equation}

 At a flip bifurcation, a fixed point loses stability and a 2-cycle is created nearby.  At bifurcations given by~\eqref{eq:flip2}, the flip bifurcation is supercritical, and a stable 2-cycle is created with zero imaginary part and real part given by
\begin{equation}
C_{l,\pm} = \pm \sqrt{\cH - \frac{1}{4} + \frac{\alpha^2}{(2l+1)^2\pi^2}}.
\label{eq:period2}
\end{equation}
This 2-cycle is stable for 
\begin{equation}
\label{eq:period2-4}
 \frac{1}{4} - \left(\frac{\alpha}{(2l+1)\pi}\right)^2 < \cH <
\frac{1}{4} - \left(\frac{\alpha}{(2l+1)\pi}\right)^2 + \frac{4 \alpha^4}{(2l+1)^6 \pi^6}.
\end{equation}
At this point, a stable 4-cycle appears.  While we have not calculated further, we conjecture that a (Hamiltonian) period-doubling cascade occurs along each branch, leading to chaotic solutions as $\cH$ is increased further.

\begin{figure}
\begin{center}
\includegraphics[width=2.5in]{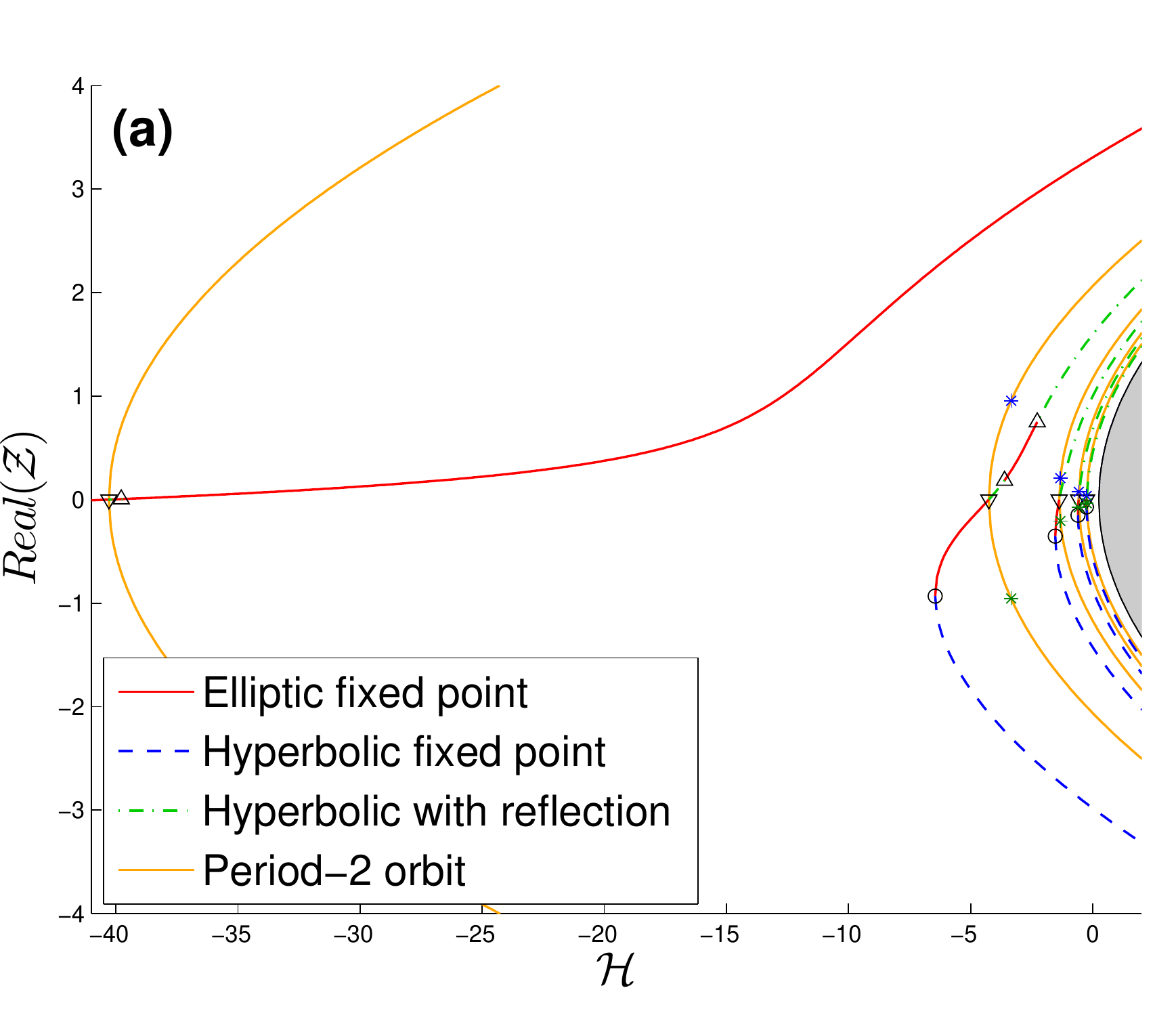}
\includegraphics[width=2.5in]{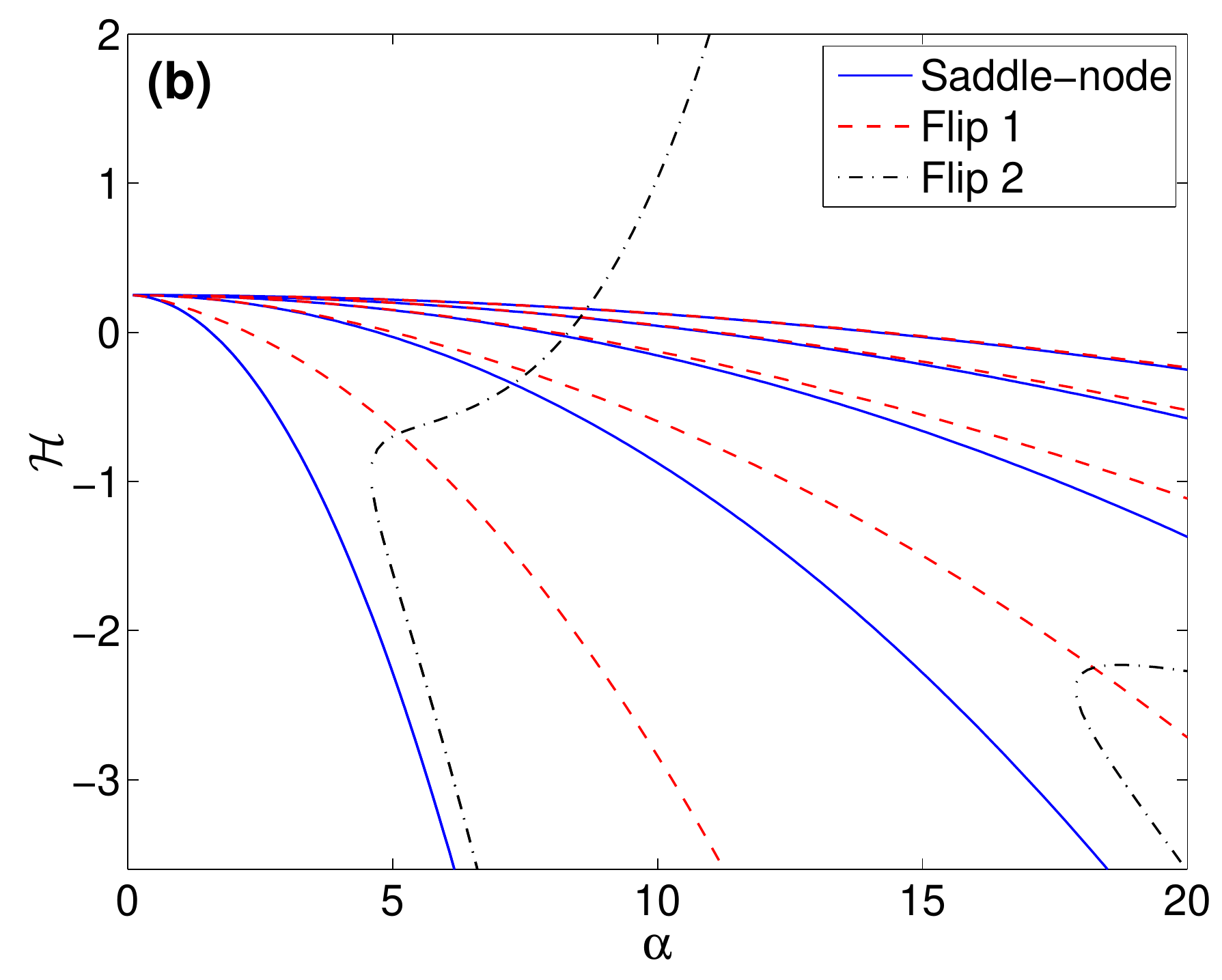}
\caption{(Color online) \subfig{a} The first five branches on a bifurcation diagram for map~\eqref{eq:morsemap} with $\alpha=20$. The bifurcations are marked: $\circ$, saddle-node~\eqref{eq:saddlenode}; $\bigtriangledown$, flip bifurcations of~\eqref{eq:flip1} $\bigtriangleup$, flip bifurcations of~\eqref{eq:flip2}(these bifurcate into the complex $\cZ$-plane, while all the branches shown lie on the real line). The black line is $\cH = C^2 +1/4$, the edge of the forbidden region $\cD_{\rm out}$ (in gray). The marked points on the period-2 curves indicate secondary period-doublings of~\eqref{eq:period2-4}. \subfig{b} A bifurcation diagram in both parameters, $\alpha$ and $\cH$.  The label `Flip 1' refers to  period-doubling bifurcations of type~\eqref{eq:flip1}, and `Flip 2' refers to those of type~\eqref{eq:flip2}. Only the first five branches of saddle node and `Flip 1' bifurcations are shown; subsequent branches approach the horizontal line $\cH = 1/4$ for fixed $\alpha$. }
\label{fig:bifurcate}
\end{center}
\end{figure}

\begin{figure}
\begin{center}
\includegraphics[width=2.5in]{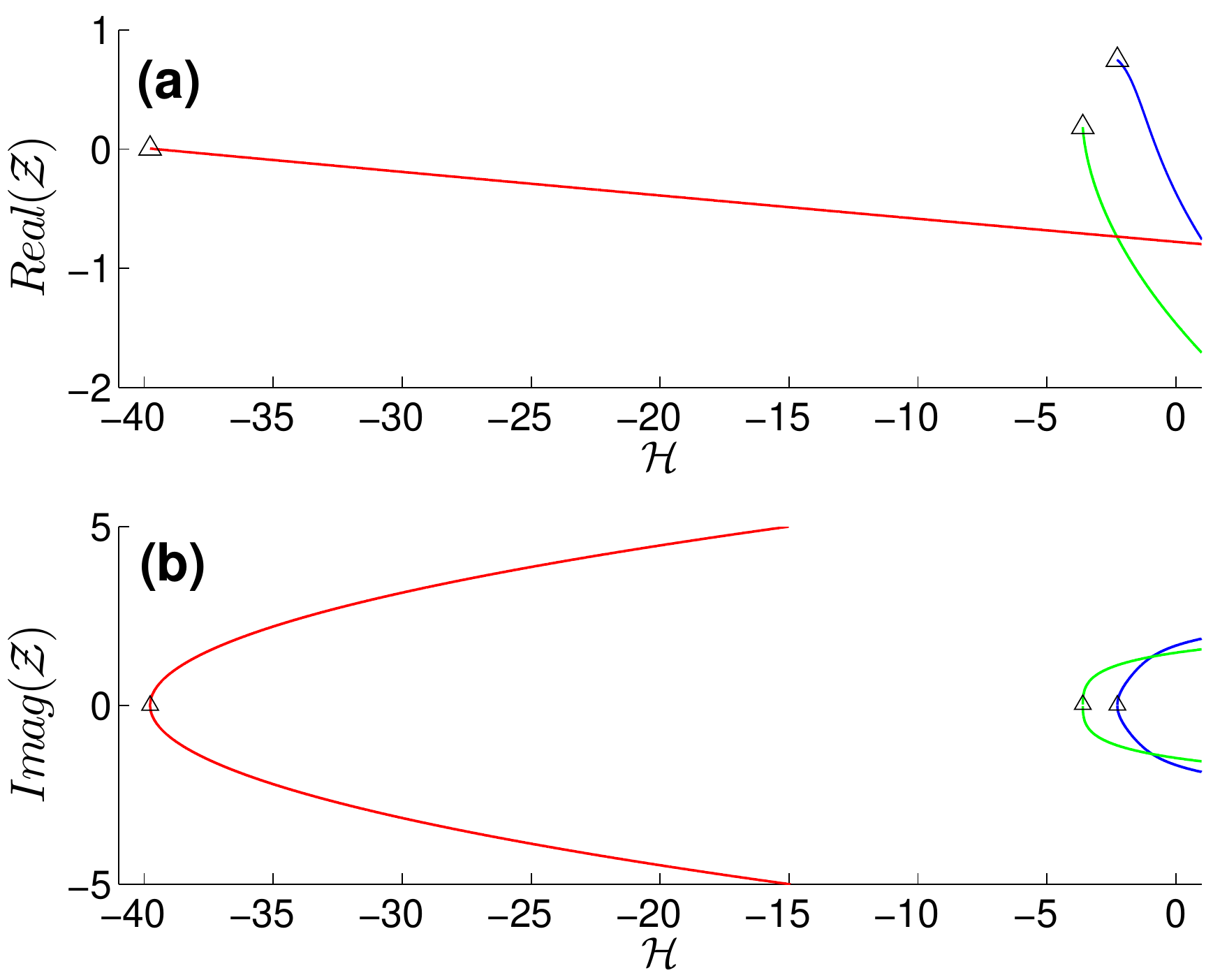}
\includegraphics[width=2.5in]{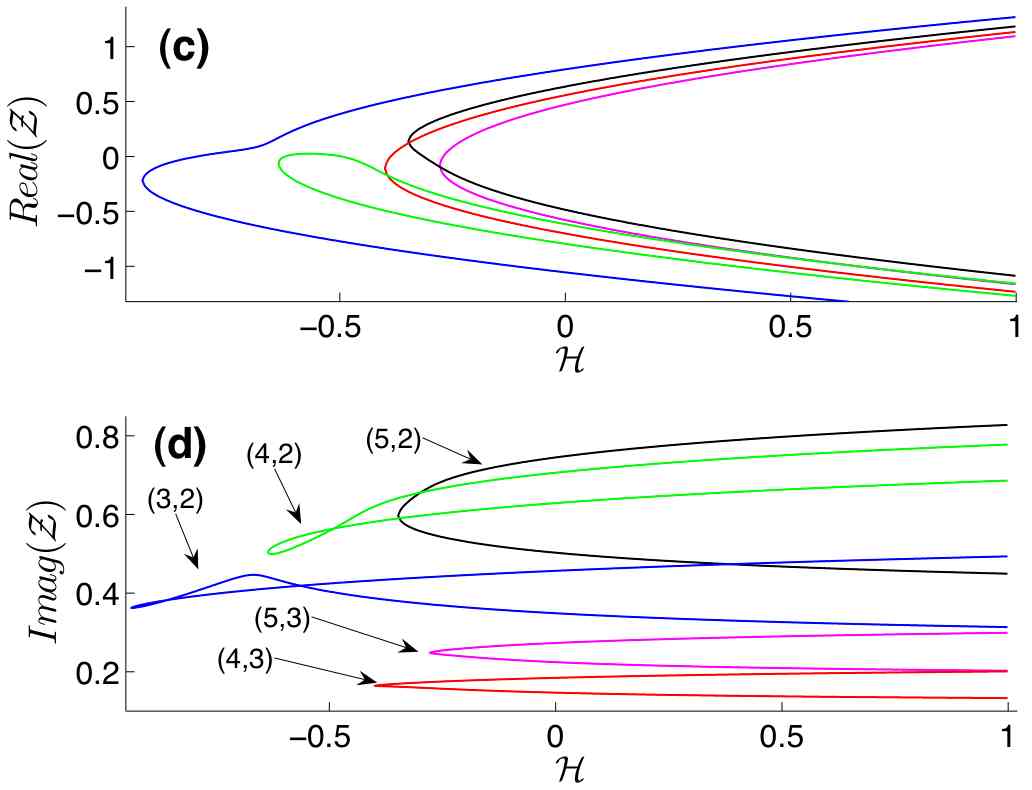}
\caption{(Color online) Bifurcation diagram for period-2 solutions of equation~\eqref{eq:period2_asym}. \subfig{a}, \subfig{b}: the real and imaginary parts of three branches that arise from the period-doubling bifurcations marked with $\bigtriangleup$ in figure~\ref{fig:bifurcate} with $\alpha=20$. \subfig{c}, subfig{d}: the first 5 branches of period-two points to bifurcate due to saddle-node bifurcations of~\eqref{eq:period2_asym}.  A symmetric branch exists reflected over the real axis.  The ordered pairs represent the numbers $(l,m)$ in equation~\eqref{eq:period2_asym}.}
\label{fig:period2bif}
\end{center}
\end{figure}

This is summarized in a bifurcation diagram showing the first few branches of $C_l$ and $C_{l,\pm}$ for $\alpha=20$ and $\cH$ varying, figure~\ref{fig:bifurcate}a, and in a bifurcation diagram over both parameters in figure~\ref{fig:bifurcate}b. An infinite sequence of branches accumulates along the edge of the region $\cH=C^2-\frac{1}{4}$.  The saddle node bifurcations approach the line $C=0$ as the branch index $l\to\infty$, so that the range of parameters for which a fixed point is (neutrally) stable on a large-$l$ branch is very small.
The 2-cycles which appear at the flip bifurcation described by~\eqref{eq:flip2} do not have a simple closed-form expression.  Instead these 2-cycles $\cF(\cZ_+) =\cZ_-$ and $\cF(\cZ_-) =\cZ_+$ with $\cZ_- = \cZ_+^*$, the complex conjugate.  Further $\cZ_\pm = C \pm i S$ satisfies
\begin{equation}
\begin{split}
\frac{C}{S-\frac{1}{2}} &= 
  \cot{\left( \frac{\alpha}{2\sqrt{C^2 + (S -\frac{1}{2})^2 - \cH}} - l\pi \right)}\\
\frac{C}{S+\frac{1}{2}} &= 
-\cot{\left( \frac{\alpha}{2\sqrt{C^2 + (S +\frac{1}{2})^2 - \cH}} - m\pi \right)},
\end{split}
\label{eq:period2_asym}
\end{equation}
where $0<l,m<\pi$.
A two-cycle resulting from the flip bifurcation~\eqref{eq:flip2} must, by continuity, satisfy $l=m$ in~\eqref{eq:period2_asym}.  Solutions to~\eqref{eq:period2_asym} may also arise in saddle-node bifurcations of this condition directly.  In this case, they may satisfy $l\neq m$, as seen in figure~\ref{fig:period2bif}.

\subsubsection*{Comparison with ODE Bifurction Diagram}

We now compare the partial bifurcation diagram of figure~\ref{fig:bifurcate}(a) with a partial bifurcation diagram of the underlying ODE~\eqref{eq:morse}.  This will help us to interpret the map's dynamics and understand their domain of validity  as an approximation to those of the ODE system. We first describe the relation between fixed points of the map~\eqref{eq:morsemap} and periodic orbits of equation~\eqref{eq:morse}.  Fixed points $\cZ_f=C_l$ of map~\eqref{eq:morsemap} are shown preceding equation~\eqref{eq:fixpt} to be real-valued.  This implies, setting $t_j=0$ in~\eqref{A_asymptote}, that as $t \to -\infty$, in the inner approximation, 
$$
A(t) \sim \cC\bigl(C_l \cos{\w t} + \frac{1}{2} \sin{\w t}\bigr).
$$
Evaluating this approximation to $A(t)$ at $t=-T_l/2$, half of the period defined by equation~\eqref{eq:theta}, we find that 
\begin{equation}
\binom{a}{\dot a} = \cC \begin{pmatrix} 
\cos{\frac{\w T_l}{2}} & -\sin{\frac{\w T_l}{2}} \\
\sin{\frac{\w T_l}{2}} & \cos{\frac{\w T_l}{2}}
\end{pmatrix}
\binom{C_l}{\frac{1}{2}}.
\label{aCS}
\end{equation}
Using the fixed-point equation~\eqref{eq:fixpt} shows that---to this order in our approximation---$\dot a(-T_l/2)=0$.  Since $\dot X(-T_l/2)=0$, this implies that the fixed points described by~\eqref{eq:fixpt} correspond to solutions of~\eqref{eq:morse} even in $t$.  Inverting the linear equation~\eqref{aCS}, 
\begin{equation}
\label{Cla}
C_l = \frac{\cos{\w T_l/2}}{\cC} a(-T_l/2).
\end{equation}

\begin{figure}[t]
\begin{center}
\includegraphics[height=2in]{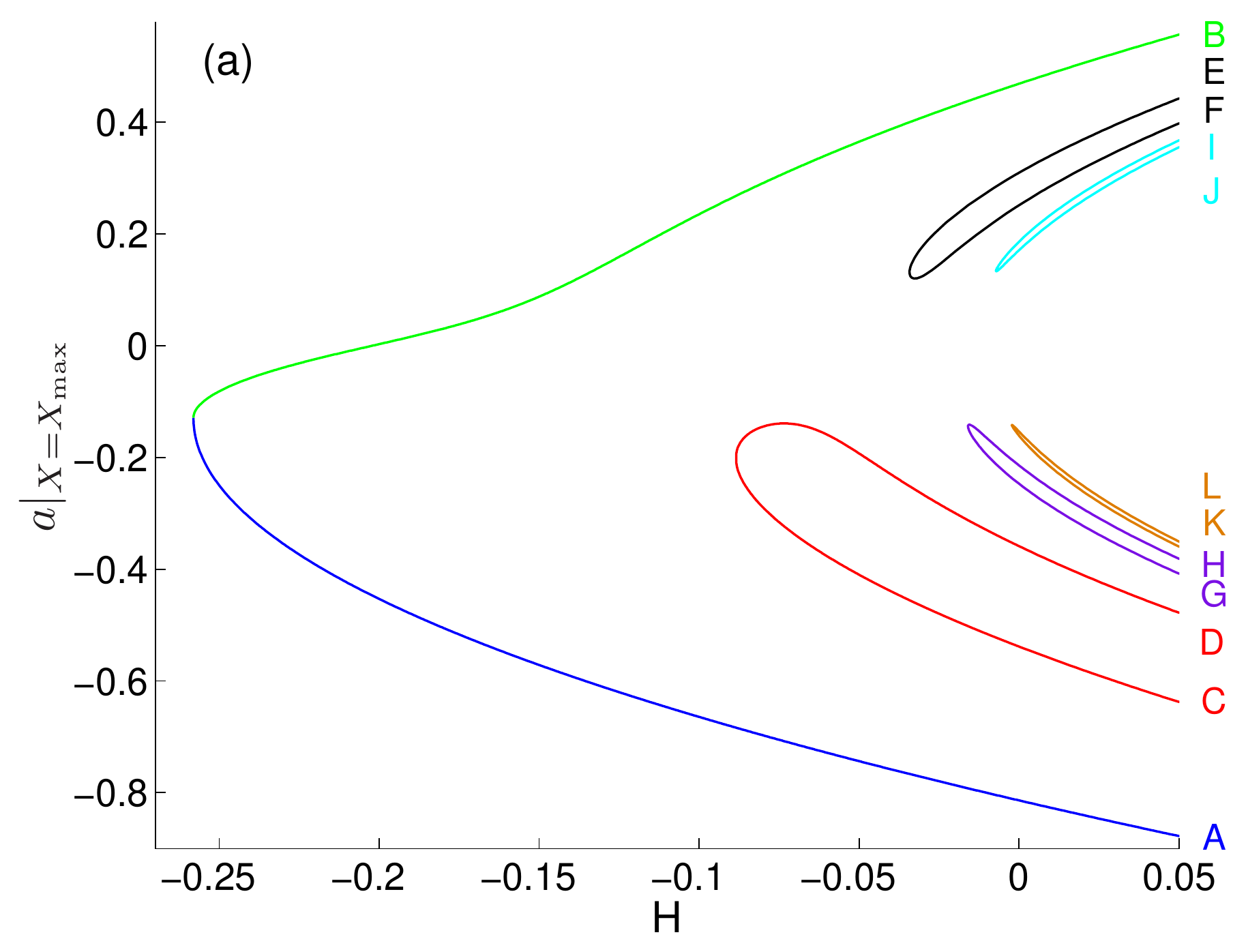}
\includegraphics[height=2in]{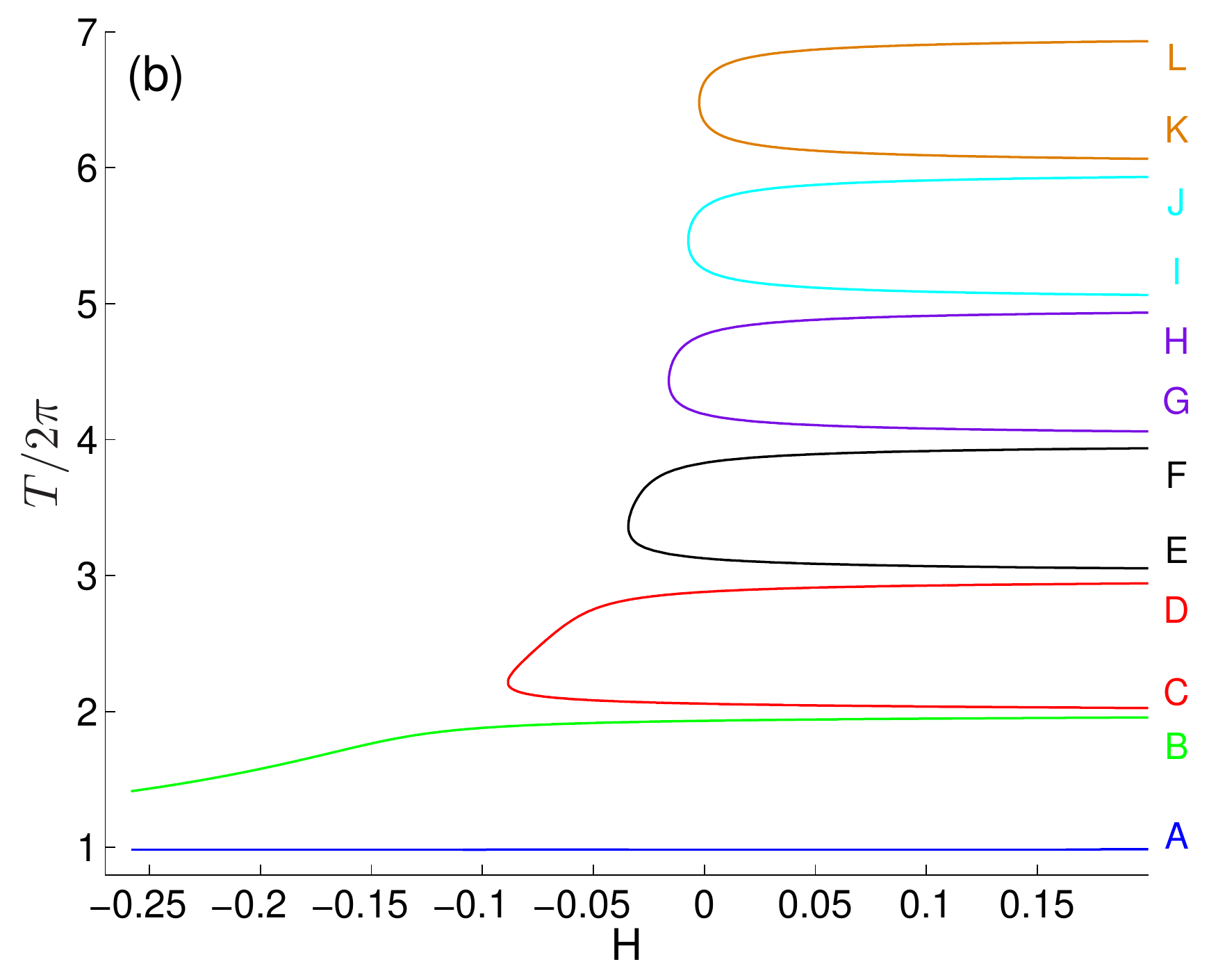}
\includegraphics[height=2in]{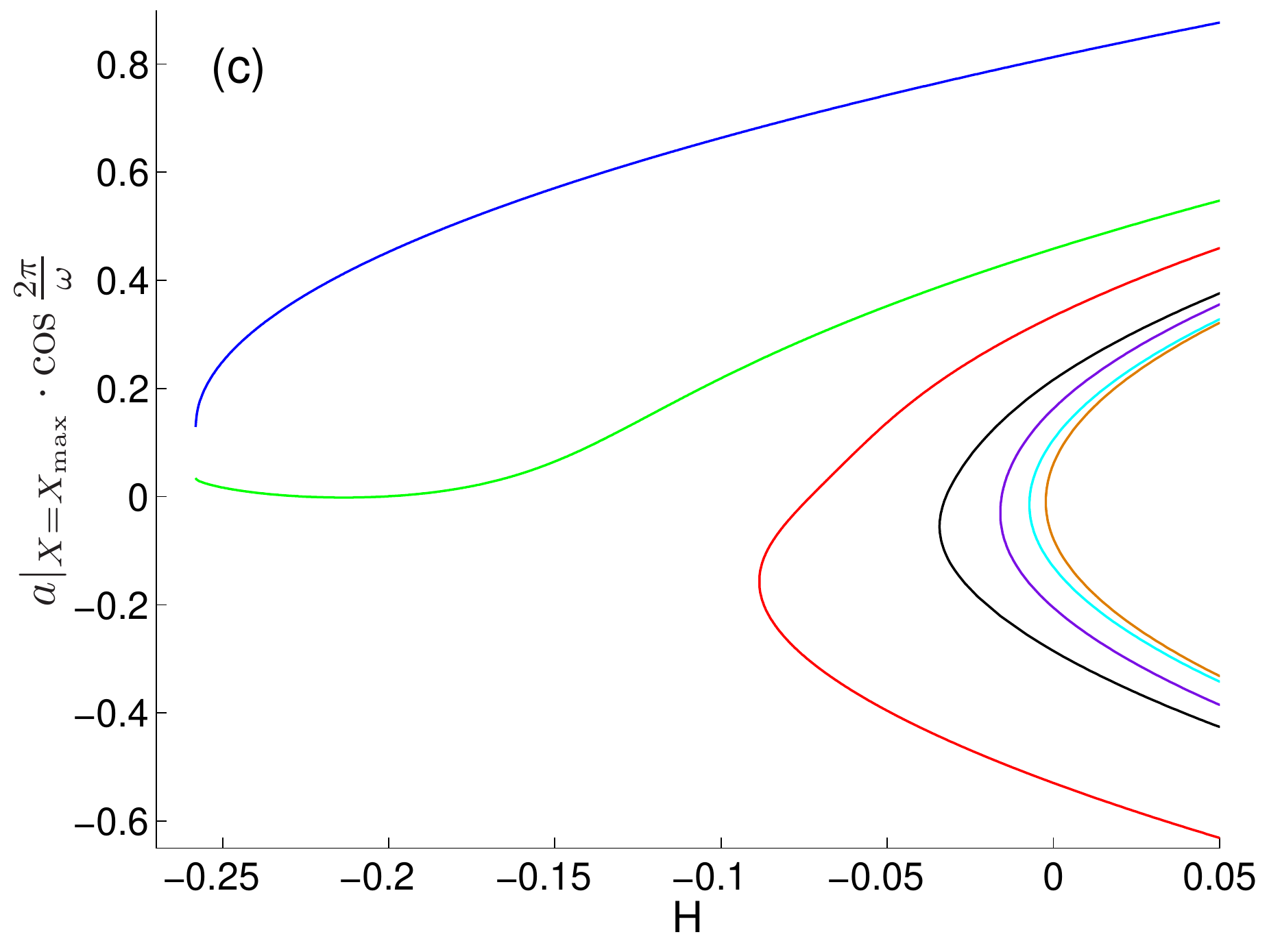}
\includegraphics[height=2in]{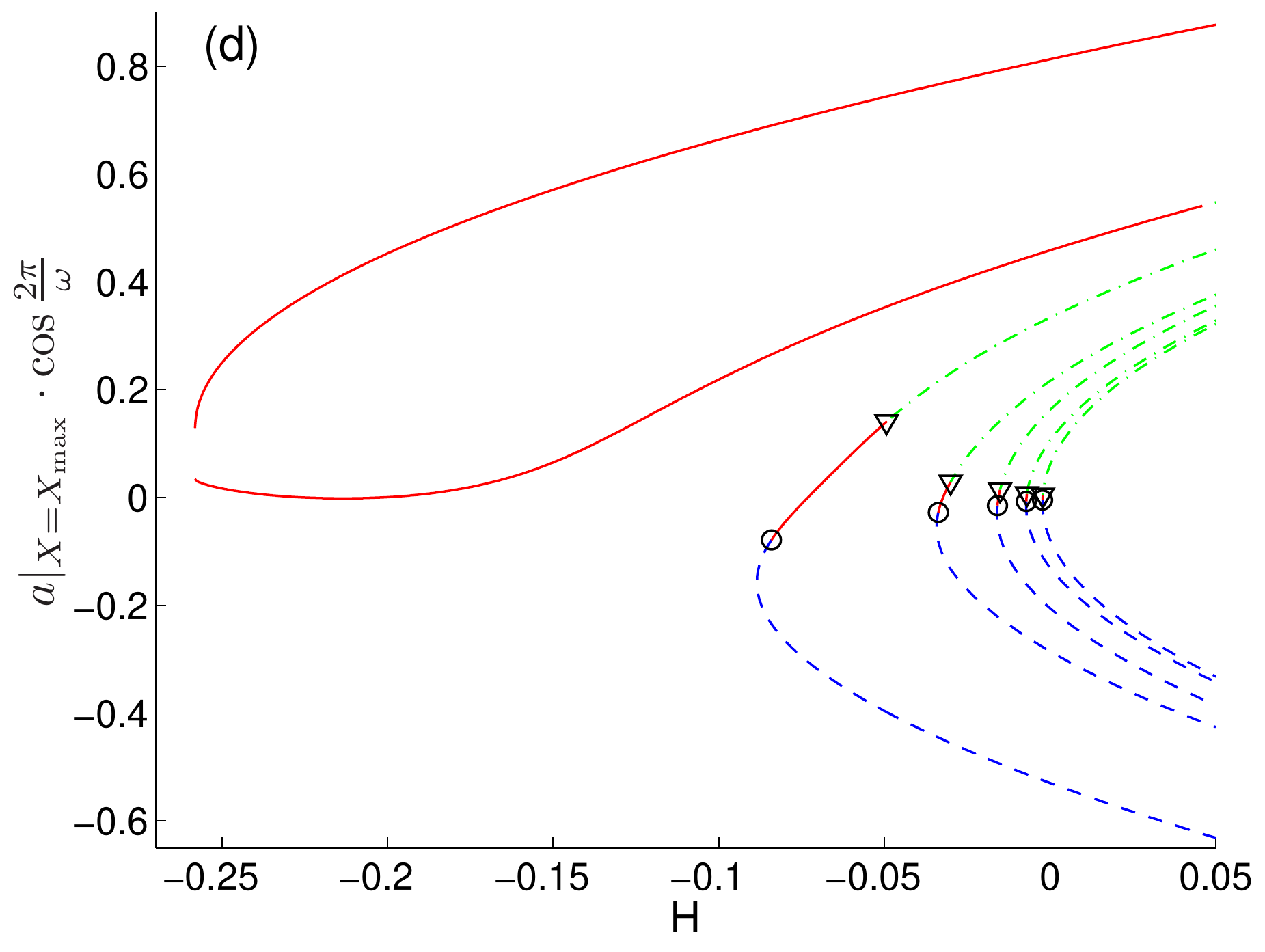}
\caption{(Color online) The seven branches of periodic orbits that cross the energy level $H=0$. Colors are used in subfigures (a), (b), and (c) to distinguish branches. \subfig{a} $a(t)$ evaluated at $X=\Xmax$. \subfig{b} The period of each branch divided by $2\pi$. \subfig{c} Approximations to $C_l$, given by equation~\eqref{Cla} to allow direct comparison to fixed points of map~\eqref{eq:morsemap}. \subfig{d} As in subfigure (c), but with the stability type indicated as in figure~\ref{fig:bifurcate}a.}
\label{fig:ode_bif}
\end{center}
\end{figure}

We numerically calculate periodic orbits of system~\eqref{eq:morse} using a method of Viswanath that combines elements of Newton's method for root-finding and the Poincar\'e-Lindstedt method for approximating closed orbits~\cite{Vis:01}.  We use this method inside a pseudo-arclength continuation algorithm to compute how the period and shape of the periodic orbit changes as a function of energy level.  Several branches of solutions are shown in figure~\ref{fig:ode_bif}.
Subfigure~\ref{fig:ode_bif}(a) shows the value taken by $a(t)$ on the periodic orbit where $X$ reaches its maximum value (related to the fixed point on the section $\dot X=0$).  There are seven branches, each plotted in a different color, five of which terminate in saddle-node bifurcations at small values of $H$.  The two branches marked A and B, corresponding to the largest values of $|a|$, merge at $H\approx-0.258$.  Subfigure (b) shows the periods of those branches. The periods of branches A and B do not approach a common value---these branches do not merge in a simple saddle-node bifurcation.    System~\ref{eq:morse} possesses a single fixed point $(X^*,a^*)=(-\log{(2-\frac{\epsilon^2}{\omega^2})},\frac{-\epsilon}{2\omega^2-\epsilon^2})$.  As $H\searrow H(X^*,a^*)=-8/31$, the periodic orbits along branches $A$ and $B$ shrink to the fixed point $(X^*,a^*)$. The frequencies of these periodic orbits approach the two different frequencies associated with the linearization about $(X^*,a^*)$.  In this sense these two branches ``end'' at $H=-8/31$---this would not be possible in a bifurcation diagram describing simple fixed points. 

The $a(t)$-component of a periodic orbit of ODE~\eqref{eq:morse} is approximately given by 
$$ a(t) \approx a_{\rm c}(t) + a_0 \cos{\w t},$$
where $a_{\rm c}(t)$ is the complementary solution due to forcing from $X(t)$ and has the same period as $X(t)$.  Along branch A, the cosine part of $a(t)$ oscillates once for each oscillation of $X(t)$, along branch B, twice.  From subfigure (a),  $a_0<0$ along branch A, while $a_0>0$ along branch B.  Along the next branch CD, $a_0<0$. At point C, the cosine part of $a(t)$ completes about two full oscillations, and as one moves along this branch toward point D an additional oscillation is added so that there are three full oscillations.  At point E, the sign of $a_0$ flips from point D, and the number of oscillations grows from three to four as one move from point D to point E, this pattern continuing from branch to branch as the period is increased.

Subfigures (c) and (d) show the quantity $C_l=a|_{X=\Xmax} \cos{\w T_l/2}$, 
defined in equation~\eqref{Cla}, which allows us to compare these calculations directly with figure~\ref{fig:bifurcate}a.  In subfigure (c), the branches are labeled using the same coloring scheme as in parts (a) and (b). Part (d) is marked as in figure~\ref{fig:bifurcate}a, with stability type calculated using the Floquet discriminant. This shows impressive agreement with that figure in some respects, but disagreement in others.  In particular, the nearly ``parabolic'' branches are nested in the same manner in both figures, and each shows the same stability behavior---a small range of $H$ for which the fixed point is neutrally stable.  The ``non-parabolic'' branch corresponding to branch B is also quite similar, likewise stable (elliptic) far into the region of positive $H$, but the final branch in the ODE figure has no analog for the discrete map.  Further, no branch extends to $H=-\infty$ as is the case for the discrete map approximation. The derivation of map~\eqref{eq:morsemap} depends on the solution staying close to the homoclinic orbit, a condition that is violated for large negative $H$ as well as along periodic orbits near the fixed point---branches A and B.

\section{Extensions}
\label{sec:extensions}
In this section, we consider extensions of the derivation of map~\eqref{eq:morsemap} in section~\ref{sec:derivation}.  The chaotic scattering phenomena described in sections~\ref{sec:intro} and~\ref{sec:prelim} have been seen in a wide variety of systems and it is worth exploring the extent to which the explanation provided by map~\eqref{eq:morsemap} and its analysis are sufficient to describe the chaotic scattering in these different settings, and to what extent the analysis needs to be modified.  We here consider two related systems.

The first  is a model for kink-defect interactions in the sine-Gordon equation~\cite{FKV:92, GH:04, GooHab:07}. It is nearly identical in form to~\eqref{eq:morse} above, with modifications to the potentials $U$ and $F$, and explicit values for the constants:
\begin{subequations}
\label{eq:sgmodel}
\begin{align}
 4 \Ddot X +  U'(X) + F'(X)a &=0; \label{eq:model_a}\\
 \Ddot a + \lambda^2 a + \e F(X) &=0 \label{eq:model_b}
\end{align}
\end{subequations}
with potentials defined by 
$$
 U(X)= -2 \sech^2(X) \text{ and }
 F(X) = -2 \tanh(X)\sech(X).
$$
and
$$\lambda^2 = \frac{2}{\e}-\frac{\e}{2}.$$

The last model we will consider was not derived in the context of solitary collisions, but is of essentially of the same form as the previous two. It was derived by Lorenz as a low-dimensional truncation of a set of ``primitive equations'' for atmospheric dynamics, and put in its present form by Camassa~\cite{Lor:86,Cam:95}. The equations take a very similar form:
\begin{equation}
\label{eq:lorenz}
\dot \psi = w- \epsilon z, \, \dot w = -R^2 \sin \psi, \, \dot x = -z, \, \dot z = x + \epsilon R^2 \sin \psi
\end{equation}
where $R$ is an order-one constant. This the canonical problem of a pendulum coupled to a linear spring.

The difference between the three systems is topological, and the topological differences between the three systems lead to differing forms for the three maps.  In both models~\eqref{eq:sgmodel} and~\eqref{eq:lorenz}, there exist two distinct separatrices as opposed to the single separatrix in model~\eqref{eq:morse}. The corresponding separatrix orbits $\Xs(t)$ are odd, and so are the coupling functions $F(X)$.  This asymmetry produces two slightly different maps depending on whether the solution is moving to the right or the left.  In addition the the $X-\dot X$ phase space of the Lorenz model~\eqref{eq:lorenz} is cylindrical.
\begin{figure}
\begin{center}
\includegraphics[width=0.3\textwidth]{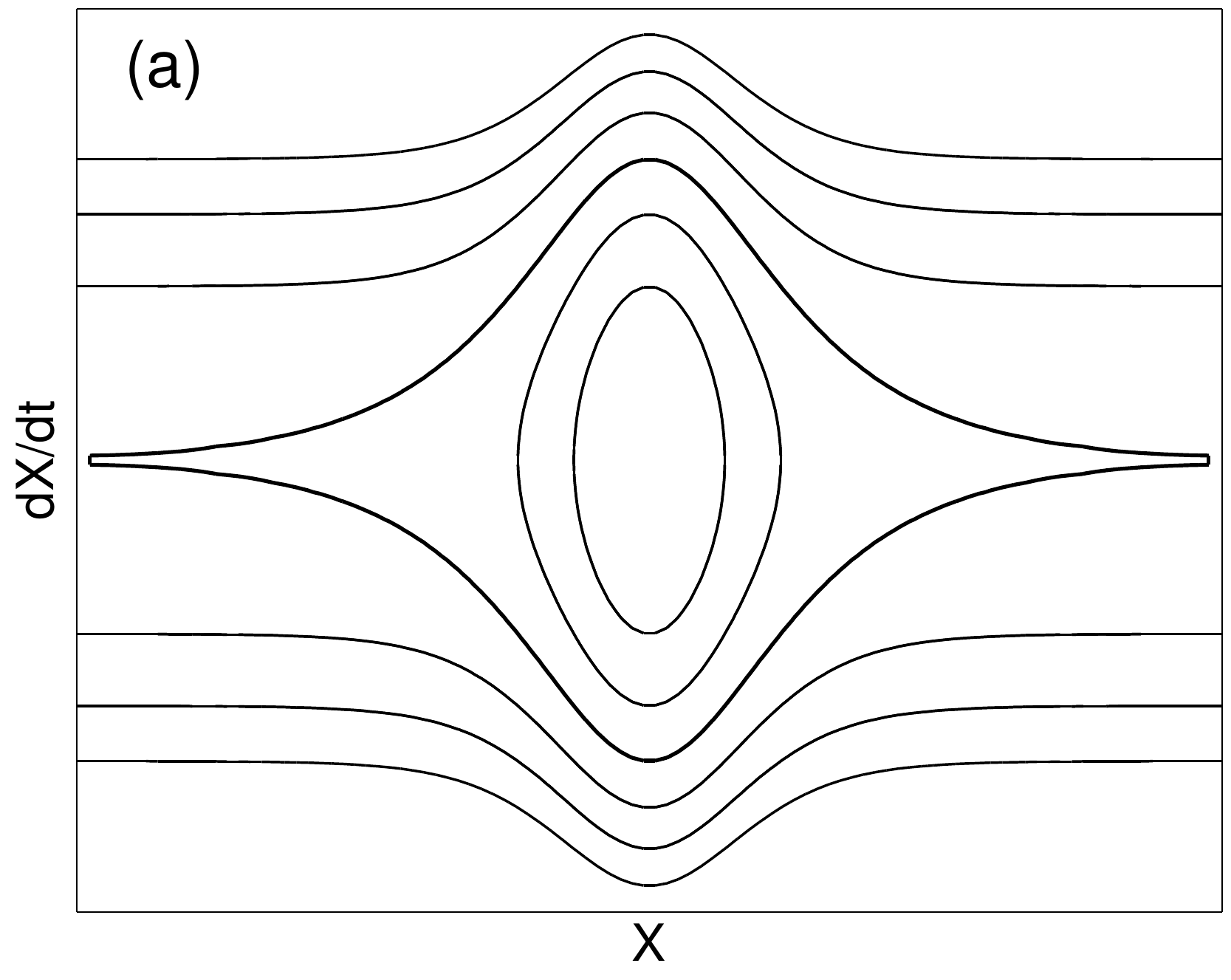}
\includegraphics[width=0.3\textwidth]{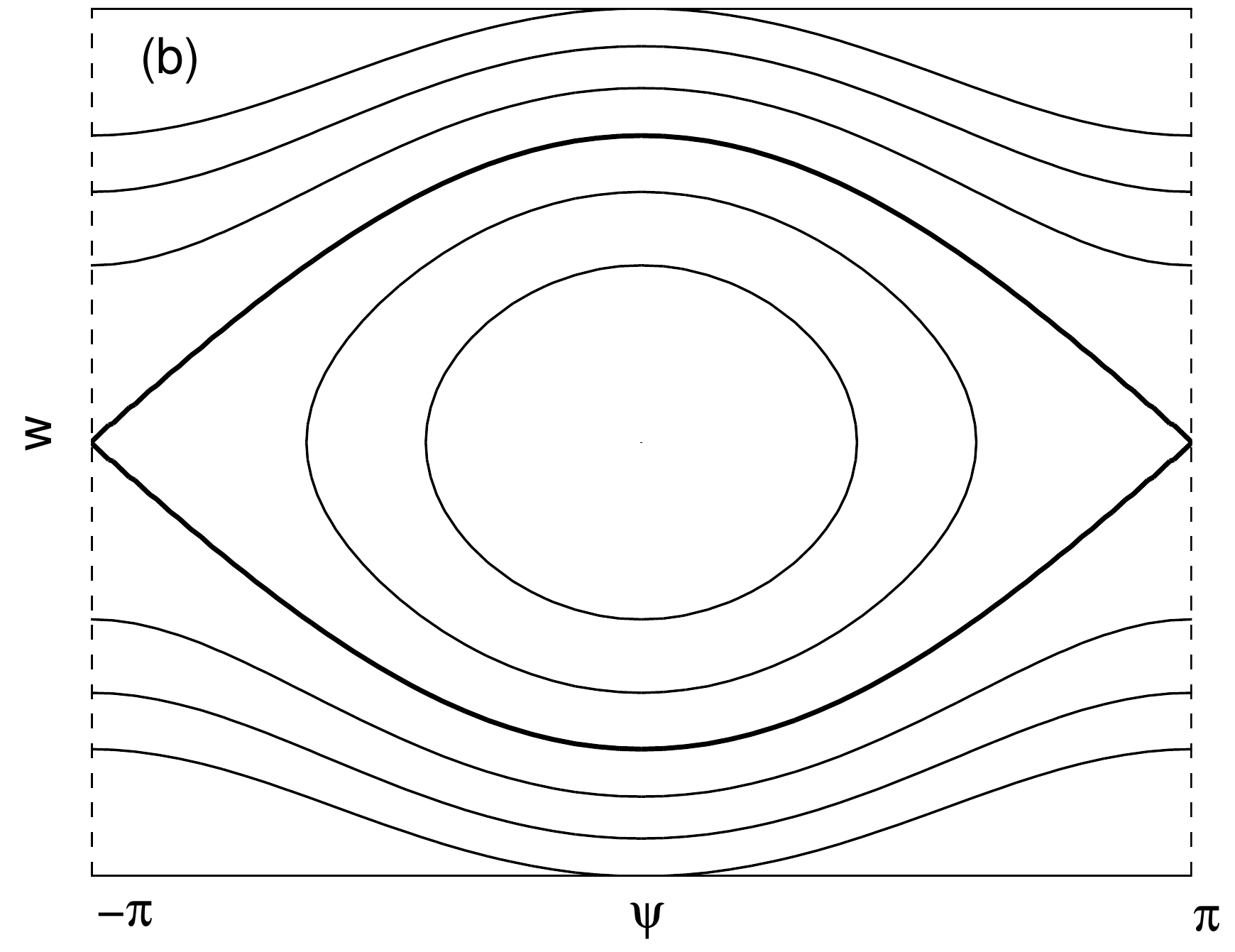}
\caption{The unperturbed phase space, with the separatrix orbit in bold, for \subfig{a} the sine-Gordon model~\eqref{eq:sgmodel}, and \subfig{b} the Lorenz model~\eqref{eq:lorenz}.  In (b), the left and right edges are identified, forming a cylinder. Compare with figure~\ref{fig:AOM}(b), which has only one separatrix.}
\label{fig:phasespace}
\end{center}
\end{figure}

\subsection{The sine-Gordon problem}
The method of deriving the map for problem~\eqref{eq:sgmodel} is identical to what has been described above.  The one difference is that the phase space for the unperturbed $X$-dynamics features two separate homoclinic orbits. $X = \pm \Xs(t) = \pm \sinh^{-1}(t)$,   Because both $\Xs(t)$ and the coupling function $F(X)$ are odd functions of their arguments, the maps obtained from the computation along the two orbits are different.  If we assume that the kink is initialized traveling to the right from $X = -\infty$, then if its initial velocity is below $\vc$, it will follow the two heteroclinic orbits in alternation until eventually enough energy is returned to the propagating mode that it can escape back to $X=\pm \infty$.

By properly scaling the variable $a(t)$, one may show that system~\eqref{eq:sgmodel} has inner and outer scalings of essentially the same form as equations~\eqref{eq:morse_outer} and~\eqref{eq:morse_inner}, allowing us to derive an iterated map in an analogous form to that in section~\ref{sec:derivation}. For solutions traveling along the upper separatrix, such that $X(t_j)=0$, and as $t\to -\infty$, 
$$
 a(t) \sim \cC\left(C_j \cos{\omega(t-t_j)} + S_j \sin{\omega(t-t_j)}\right)
$$
and $\cC$ is again, for the moment, undetermined.  Solving for $a(t)$ by variation of parameters, as in equation~\eqref{eq:A_int} and allowing $t-t_j\to\infty$, one finds
\begin{equation}\begin{split}
 a(t) \sim &\left(\cC C_j 
 +\frac{\e}{\lambda}\int_{-\infty}^{\infty} F(\Xs(\tau-t_j)) \sin{\lambda(\tau-t_j)} d\tau\right)\cos{\lambda(t-t_j)} \\
 & + \left(\cC S_j -\frac{\e}{\lambda}  \int_{-\infty}^{\infty} F(\Xs(\tau-t_j)) \cos{\lambda(\tau-t_j)} d\tau\right)\sin{\lambda(t-t_j)}.
 \label{eq:a_int}
 \end{split}\end{equation}
 Because $F(\Xs(t)$ is odd, the integral in the coefficient of $\sin{\lambda(t-t_j)}$ vanishes.  We define 
 \begin{multline*}
\cC  = \frac{\e}{\lambda}  \int_{-\infty}^{\infty} F(\Xs(\tau)) \sin{\lambda\tau} \ d\tau
=\Imag \frac{\e}{\lambda}  \int_{-\infty}^{\infty} F(\Xs(\tau)) e^{i\lambda\tau} \ d\tau \\
 =\frac{\e}{\lambda} \int_{-\infty}^{\infty} \frac{-2\tau}{1+\tau^2} e^{i\lambda \tau} \ d\tau
= \frac{-2 \pi \epsilon e^{-\lambda}}{\lambda}
\end{multline*}
using the exact form of $\Xs$ and the residue theorem.  Then 
$$
a(t)\sim\cC(C_j+1)\cos{\lambda(t-t_j)} + \cC S_j \sin{\lambda(t-t_j)}
\text{ as } t-t_j \to +\infty.
$$
As in section~\ref{sec:derivation}, we recast this as the asymptotic behavior as $t-t_{j+1} \to -\infty$ for the next interaction.  Using the same type of expansion as in that section, we find 
$t_{j+1}- t_j \sim \pi/\sqrt{-2 E_{j+1}}$.    A calculation like~\eqref{eq:DE1} shows that $\Delta E =  -2\pi^2 \epsilon e^{-2\lambda}(1-2C_j)$.
Letting, $E_j = 2\pi^2 \epsilon e^{-2\lambda} \cE_j$, the three dimensional map can be written
\begin{equation}\begin{split}
\cE_{j+1} &= \cE_j - (1-2C_j)\\
C_{j+1} &= \cos{\theta_{j+1}} (C_j-1) + \sin{\theta_{j+1}}S_j \\
S_{j+1} &= -\sin{\theta_{j+1}} (C_j-1) + \cos{\theta_{j+1}}S_j 
\end{split}
\label{eq:map_3components}
\end{equation}
where $\theta_{j+1} = \lambda (t_{j+1}-t_j) = \lambda\pi/\sqrt{-2 E_{j+1}}$.
Using the conserved energy $\cH = \cE_j + C_j^2 + S_j^2$  and again letting $Z_j = C_j + i S_j$ yields the reduced map:
\begin{equation}
\label{eq:sgmap1}
Z_{j+1} = e^{\frac{i\alpha}{\sqrt{\left\lvert Z_j -1\right\rvert^2 -\cH}}}(Z_j -1)
\end{equation}
where $\alpha  = \lambda e^{\lambda}/2\sqrt{\epsilon}$. Repeating this computation with the solution tracing along the leftgoing heteroclinic orbit, the only change to map~\eqref{eq:sgmap1} is that wherever $(Z_j -1)$ appears in the expression, it is replaced by $(Z_j+1)$. 

If we assume that the solution first traces along a rightgoing heteroclinic orbit, then the general map is:
\begin{equation}
\label{eq:sgmap}
Z_{j+1} =  e^{\frac{i\alpha}{\sqrt{\left\lvert Z_j  +(-1)^j\right\rvert^2 -\cH}}}(Z_j +(-1)^j).
\end{equation}
This map has no fixed points, but one may look for solutions that satisfy $Z_{j+2} = Z_j$ and discover that this system does have fixed point, and one finds a bifurcation diagram very similar to figure~\ref{fig:bifurcate}a.

\subsection{The Lorenz model}
Model~\eqref{eq:lorenz} differs from the first two in two important ways.  First, the evolution takes place on  a cylinder.  In the previous two models, the map became undefined if the solution to the ODE escaped to $X =\pm \infty$.  In the present model, the analogous behavior is simply an orbit that wraps around the cylinder, in which case the map can be applied again.  The second major difference is that the has genuine fixed points at $\psi = \pm \pi$, of mixed elliptic-hyperbolic type in the full four-dimensional phase space, as opposed to the degenerate fixed points at infinity displayed by the first two models.  This has a direct consequence on the general form of the  asymptotic approximation of $(t_{j+1} - t_j)$.

Solutions to~\eqref{eq:lorenz} conserve a Hamiltonian $H = \frac{w^2}{2} - R^2 (\cos\psi+1) + \frac{x^2}{2} + \frac{z^2}{2}$.  The energy in the propagating mode is $E =  \frac{w^2}{2} - R^2 (\cos\psi+1)$, where the constant has been chosen to make $E=0$ along the heteroclinic orbit.

The separatrix orbit in~\eqref{eq:lorenz} is given by $\psi_{\rm S} = \pm \sin^{-1}{\tanh {R t}}$.  Assuming that the solution follows the upper heteroclinic orbit, centered at time $t_j$, with 
$$
x \sim \cC \left( C_j \cos{(t-t_j)} + S_j \sin{(t-t_j)} \right)
$$
then 
\begin{align*}
x = &\cC(C_j \cos{(t-t_j)} + S_j \sin{(t-t_j)} + \\
&\epsilon R^2 \left( -\sin {(t-t_j)} \int_{-\infty}^t \sin \psi_{\rm S}(\tau -t_j) \cos{(\tau-t_j)} \ d\tau + 
\cos {(t-t_j)} \int_{-\infty}^t \sin \psi_{\rm S}(\tau -t_j) \sin{(\tau-t_j)} \ d\tau \right).
\end{align*}
As $t \to \infty$, the integral in the coefficient of $\sin{(t-t_j)}$ vanishes.  Once more using the exact form of $\psi_{\rm S}$ and the residue theorem, we define
$$
\cC = \epsilon R^2 \int_{-\infty}^{\infty} \sin{\left(\psi_{\rm S}(\tau -t_j)\right)} \sin{(\tau-t_j}) \ d\tau 
= \epsilon R^2 \int_{-\infty}^{\infty} 2 \sech{Rt} \tanh{Rt} \sin{t} \ dt
= 2 \pi \epsilon\sech{\frac{\pi}{2R}}
$$ 
where integration by parts and the residue theorem are used to evaluate the integral.
Then as $t-t_j \to +\infty$,
$$
x \sim \cC\bigl( (C_j+1) \cos{(t-t_j)} + S_j \sin(t-t_j) \bigr).
$$
Now,  assuming that the near-saddle approach occurs on the level set of $E_{j+1}$, we find
$$
t_{j+1}-t_j = \frac{2}{R}\log{\frac{4\sqrt{2}R}{\sqrt{\lvert E_{j+1} \rvert}}}.
$$
This is defined regardless of the sign of $E_{j+1}$, in contrast to the first two examples. 
This is a well known result describing the passage time near a saddle point and appears, for example, in Lichtenberg and Lieberman's textbook~\cite{LL}. The rate at which $\Delta t$ diverges as $E_j \to 0$ is logarithmic, as opposed to the previous examples in which $\Delta t \propto E^{-1/2}$.  An equation like~\eqref{eq:Mel1}, yields
$$E_{j+1} = E_j  -2 \pi^2 \epsilon\sech^2{\frac{\pi}{2R}}(1+2C_j).$$
Making the scaling $E_j = 2 \pi^2 \epsilon\sech^2{\frac{\pi}{2R}} \cE_j$, defining $Z_j = C_j + i S_j$, and using the conserved quantity  $\cH = \cE_j + C_j^2 + S_j^2$, the map can be written as 
$$
Z_{j+1} = e^{i \frac{2}{R} \log \left(\frac{4\sqrt{2}R}{\cC\sqrt{\left|\cH-|Z_j+1|^2\right|}}\right)} (Z_j +1)
$$
which is defined as long as $|Z_j+1|^2 \neq \cH$.

If the solution traverses the lower separatrix orbit $\psi \approx -\psi_{\rm S}(t-t_j)$, the above formula is modified by replacing $(Z_j+1)$ with $(Z_j-1)$. The maps are not defined in strict alternation, as for~\eqref{eq:sgmap}---which formula applies is determined by whether the pendulum is swinging to the right or to the left at a given step.  We can keep track of this by introducing a discrete variable $P_j$ that can take only the values $P_j = \pm 1$.  The state of the system can be described by the variables $Z_j$ and $P_j$ where 
\begin{subequations}
\label{eq:lorenzmap}
\begin{align}
Z_{j+1} & = e^{i \frac{2}{R} \log \left(\frac{4\sqrt{2}R}{\cC\sqrt{\left|\cH-|Z_j+P_j|^2\right|}}\right)} (Z_j +P_j);
\label{eq:lorenz_Z}\\
P_{j+1} & = \sign{\left(\cH - |Z_j+P_j|^2\right)} P_j.\label{eq:lorenz_P}
\end{align}
\end{subequations}
The rule for $P_j$ says that if the energy $\cE_{j+1}  = \cH - |Z_j+P_j|^2$ is positive, then on the next pass, the pendulum will continue swinging in the same direction, but if it is negative, it will turn around and swing in the opposite direction on the next pass.

This can be thought of as an iterated-map analog of a hybrid dynamical system.  A formal definition of hybrid dynamical systems is given by Guckenheimer and Johnson~\cite{GucJoh:95}.  Briefly, a hybrid dynamical system consists of an indexed collection of one or more ordinary differential equations.   An initial condition specifies both which equation is to be solved as well as what the initial values to be used for its solution.  The solution is integrated until some time $t_1$ when the trajectory crosses some specified manifold (if it ever does).  At this point a discrete map is applied that tells the system which ODE to solve next and what initial conditions to use.  This process then continues, producing a sequence of transition times $t_j$ and trajectories defined on intervals $(t_{j-1},t_j)$.  Map~\eqref{eq:lorenzmap} is the discrete time analog.  At each step, one must determine the value of the parameter $P_j$ in order to determine whether the pendulum is swinging clockwise or counterclockwise, and thus which form of the map to apply.   The two maps are continuous (except on a singularity surface) maps, and a given map is iterated until it produces a negative value of $\cE_{j+1}$, at which point, the evolution continues by iterating the other map.  We do not know of any other examples of this type of map that have been studied.

In the work of Camassa et al., the focus was on showing the existence of multipulse homoclinic orbits~\cite{Cam:95,CamKovTin:98}.  Such homoclinic orbits take place on the energy level $\cH=0$ so that $P_{j+1}=-P_j$ and the novel aspects of the map are missing.  The methods described in section~\ref{sec:nbounce} make finding multipulse homoclinics much easier.  When $\cH>0$, the map becomes significantly more interesting, and equation~\eqref{eq:lorenzmap} may be used to locate periodic orbits which may move clockwise and counterclockwise an arbitrary number of times in any sequence.

\section{Discussion and future directions}
\label{sec:conclusion}
In this paper we have studied an iterated-map model of solitary wave collisions.  Our previous studies used a similar analytic method, but produced an iterated map that was significantly harder to analyze and which applied to only a limited set of initial conditions~\cite{GH:04,GooHab:05,GH:05,GooHab:07}.  This paper analyzed in depth the map generated as an approximation to one such system. Chaotic scattering of a very similar type has been seen in many solitary wave collisions, and the methods used to analyze this map will not differ greatly, nor we believe, will the types of structures found, for example, in the bifurcation diagrams for such maps as we derived in section~\ref{sec:extensions}.  

We briefly discuss a few avenues not explored in the present study but of interest for future papers.

\subsubsection*{Modeling of dissipation}

A key difference between the PDE and ODE simulations is the possibility of true capture for all time, as seen by comparing figures~\ref{fig:phi4v} and~\ref{fig:ode_in_out}.  By conservation of phase-space area, almost all solutions to equation~\eqref{eq:morse} that approach $X=+\infty$ as $t\to-\infty$ will eventually escape as $t\to+\infty$.  By contrast, the loss of kinetic energy to radiation acts as a damping to equation~\eqref{eq:phi4} and makes capture possible for a significant set of initial conditions.

One may modify equation~\eqref{eq:AOM} or~\eqref{eq:morse} to include the effects of this dissipation.  Such a dissipative model is derived perturbatively for the interaction of sine-Gordon kinks with localized defects in~\cite{GHW:02}.  The effect was to modify the equivalent of equation~\eqref{eq:A_morse} to something similar to
$$
\ddot A + \e^3 A^2 G(X) \dot A + \w^2 A + \epsilon F(X) =0,
$$
where $G(X)>0$ and $G(X)\to 0$ as $X\to \infty$, i.e.\ the damping term is nonlinear in $A$ and decays away from $X=0$, so that damping only occurs when the kink and antikink are near each other.  This should have the effect of modifying the equation for $\cE_j$ in map~\eqref{eq:twocomponent} and destroying the conservation law~\eqref{eq:H}.  Thus,  writing $\cH_j= \cE_j+ \abs{Z_j}^2$, we find $\cH_j$ is strictly decreasing away from any fixed points.  As $\cH_j$ decreases, the escape regions $\cDo$ shrink to a point and then disappear.  If the disk disappears without the solution ever landing on it, then the solitary waves will be trapped forever.  A detailed analysis of this dynamics is a topic for later research.  Numerical simulations of the dissipative system in the above reference shows that this mechanism is enough to remove many but not all of the resonance windows.  A systematic study of such modified maps is underway.

\subsubsection*{Invariant manifold calculations}

In order to understand the fractal dynamics described in this paper, it is necessary to understand not merely the fixed points of map~\eqref{eq:morsemap} but their stable and unstable manifolds, which control the topological organization of the phase space.  The subject of lobe dynamics, first developed by Rom-Kedar to understand the topology of such stable manifolds~\cite{Rom:90,Rom:94} have the potential to be of great use.  First of all, using measure-theoretic extensions of these methods due to Meiss, should be useful for calculating quantities such as the average number of times two solitons collide before escaping~\cite{Mei:97}.  Recent work by Mitchell and Delos~\cite{MitDel:06} allows one to gain a detailed understanding of the topology of the invariant manifolds and may be useful for gaining a more complete picture of the interleaving of the different scattering trajectories, far beyond the simple Cantor-set descriptions of section~\ref{sec:sitnikov}.

This analysis is complicated in two ways by the infinite winding around the disks $\cDi$ and $\cDo$ in map~\eqref{eq:morsemap} with $\cH>0$.   First, it implies the existence of an infinite number of fixed points, whose invariant manifolds may be important to the complete application of the above methods.  Secondly, the invariant manifolds which cross these disks are split into an infinite number of disconnected pieces, so standard methods, which assume that these manifolds are continuous, are not directly applicable.  It may be possible to avoid this difficulty by using different types of iterated-map reductions than equation~\eqref{eq:morsemap}, or it may be necessary to modify the methods in order to make them directly applicable to this map.

\subsubsection*{Application of methods directly to PDE simulations}
One of the applications often discussed for solitary wave-defect collisions is to build optical components such as switches or logic gates with them.  Without getting into details, it would be necessary to understand figures such as figure~\ref{fig:phi4_excited} for those particular systems in order to tune initial conditions to be sufficiently far away from the boundaries between the scattering regions.  To accomplish this, it should be necessary to implement many of the ideas of this paper directly to  Poincar\'e maps built directly from numerical simulations of partial differential equations.

\section*{Acknowledgements}
It is a pleasure to thank Richard Haberman, for his earlier collaboration on this topic and careful reading of this manuscript,  Philip Holmes, who pointed me to the Sitnikov three-body problem, Divakar Viswanath, for assistance in using his numerical method, and Kevin Mitchell for comments that led to the computations in figure~\ref{fig:ode_bif}.  The author received support from NSF grant DMS-0506495 and performed some computations on equipment provided by NSF DMS-040590.

\providecommand{\bysame}{\leavevmode\hbox to3em{\hrulefill}\thinspace}
\providecommand{\MR}{\relax\ifhmode\unskip\space\fi MR }
\providecommand{\MRhref}[2]{%
  \href{http://www.ams.org/mathscinet-getitem?mr=#1}{#2}
}
\providecommand{\href}[2]{#2}

\end{document}